\documentclass{aastex}          %% The manuscript based on AASTeX v5.x
\usepackage{spr-astr-addons}    %% mimicing ASTR journal style
                              
\usepackage{url}
\usepackage{txfonts}
\usepackage{graphicx}
\usepackage{pdflscape}
\usepackage{color}

\usepackage{hyperref}
\begin{document}
%% Article title
%
\title{Characterizing Some Gaia Alerts with LAMOST and SDSS}

%% Running heads
\shorttitle{Characterizing  Gaia Alerts with LAMOST and SDSS}
\shortauthors{Z. Huo et al.}

%% Author and Affilations
\author{Z. Huo \altaffilmark{1}} 
\and 
\author{M. Dennefeld\altaffilmark{2}}%\thanks{This work was initiated during a visit to SWIFAR (in the frame of the 
%Chinese-French LIA "Origins" program), whose hospitality and support are gratefully acknowledged. }
\and
\author{X. Liu \altaffilmark{3}}
\and
\author{T. Pursimo \altaffilmark{4}}
\author{T. Zhang \altaffilmark{1,5}}
%\affil{}
\email{} %% non-output

%% Alternate Affilations
\altaffiltext{1}{CAS Key Laboratory of Optical Astronomy, National Astronomical Observatories, Chinese Academy of Sciences, 10101, Beijing;   
              \it{Email: zhiyinghuo@nao.cas.cn} }
\altaffiltext{2}{Institute of Astrophysics Paris (IAP), and Sorbonne University, 98bis Boulevard Arago, F-75014 Paris;
             \it{Email: dennefeld@iap.fr}}
\altaffiltext{3}{South-Western Institute For Astronomy Research (SWIFAR), Yunnan University, Kunming, Yunnan 650091, P.R. China }
\altaffiltext{4}{Nordic Optical Telescope (NOT), Apartado 474, E-38700 Santa Cruz de La Palma, Spain }
\altaffiltext{5}{School of Astronomy and Space Science, University of Chinese Academy of Sciences, 101408, Beijing}

% Abstract
\begin{abstract}
 The ESA-Gaia satellite is regularly producing Alerts on objects where
   photometric variability has been detected after several passages over the 
   same region of the sky. 
  The physical nature of these objects has often to be determined with the help of complementary observations 
    from ground-based facilities. 
   We have compared the list of Gaia Alerts (from the beginning in 2014 to Nov. 1st, 2018) with 
   archival LAMOST and SDSS spectroscopic data. 
   A search radius of 3$\arcsec$ has been adopted. 
   In using survey data, the date of the ground-based observation rarely corresponds to the date of the Alert, 
   but this allows at least the identification of the source if it is persistent, or the host galaxy if the object 
   was only transient like a supernova (SN). Some of the objects have several LAMOST 
  observations, and we complemented this 
   search by adding also SDSS DR15 data in order to look for long-term variability. A list of Gaia  Nuclear Transients 
   (GNT) from \cite{kostrz18} has been included in this search also. 
   We found 26 Gaia Alerts with spectra in LAMOST+SDSS labelled as stars, among which 12 have multi-epoch spectra. 
   A  majority of them are Cataclysmic Variables (CVs). 
   Similarly, 206 Gaia Alerts have associated spectra labelled as galaxies, among which 49 have multi-epoch spectra. Those spectra were generally obtained on a date widely different from  the Alert date, and are mostly emission-line galaxies with no particularity (except a few Seyferts), leading to the suspicion that most of the Alerts were  due to a SN. 
   As for the GNT list, we found 55 associated spectra labelled as galaxies, among them 13 with multi-epoch spectra. 
   In these two galaxy samples, in only two cases, Gaia17aal and GNTJ170213+2543,  was the date of the spectroscopic observation   
    close enough to the Alert date:  we find a trace of the SN itself in their LAMOST spectrum, both being now classified here as a type Ia SN.  Compared to the galaxy sample from the Gaia alerts, the GNT sample has a higher proportion of  AGNs, suggesting that some of the detected    variations are also due to the AGN itself. 
  Similarly for Quasars, we found only 30 Gaia Alerts but  68 GNT cases associated with single epoch quasar spectra in the databases. 
   In addition to those,  12 plus 23  are  quasars where multi-epoch spectra are available. 
   For ten out of these 35, their multi-epoch spectra  show appearance or disappearance of the broad 
    Balmer lines and also variations in the continuum,  qualifying them as ``Changing Look Quasars" and therefore significantly increasing the available sample of such objects. 
\end{abstract}

% Keywords

\section{Introduction}

 A few months after its launch in December 2013, the ESA-Gaia mission (Gaia Collaboration \citeyear{gaia16}) 
 started to produce regularly  Alerts on photometric variability of targets which crossed its field of view \footnote{http://gsaweb.ast.cam.ac.uk/alerts}. 
The variability is derived from comparison of actual photometry  with values measured during previous passages, without any 
 reference to magnitudes available in external catalogues. This procedure  avoids the problem of having to 
compare magnitudes in various different systems, with all the associated calibration issues, but
does not lead to an immediate identification of the variable object. A note is  provided with the Alert on possible associations with 
catalogued objects, but is based on positional proximity only. Some 
complementary information is available through the very low dispersion spectra  (R $\sim 30$)
obtained on board by the Blue and the Red Photometers \citep[BP and RP,][]{gaia16},
 but this low resolution does not always allow an unambiguous classification. 
\cite{blago14} have made simulations showing that about 75$\%$ of the transients should be 
robustly classified by BP/RP,  but in practice this classifier requires to be trained first with secured classifications. After a few years 
of experience, which helped to improve the internal classification, it appears for example  that type Ia SNe are rather reliably identified, 
but other types of objects much less so. 
As a result, many objects still have the qualification of  ``Unknown" in the table of Alerts, particularly at the beginning of the survey.  
The light curve is another, important  element for classification, but the irregular sampling with Gaia itself means 
that a long period of time is necessary before a proper light curve is obtained. 
It is therefore important to obtain complementary ground-based data  to identify the exact nature of each variable source. \\

 We have therefore started to search for associations between objects appearing in the Gaia Alerts and 
 spectra obtained with the LAMOST Survey \citep{Cui2012, Luo2015} and/or with the Sloan Digital Sky Survey 
\citep[SDSS-DR15;][]{Aguado2019}. 
In addition to those Gaia Alerts, we have also included in our search the Gaia Nuclear Transients (GNT) 
 detected by a different method by \cite{kostrz18}.  
This paper presents the spectral information  retrieved during our search. In Section 2, we present the search of associations and the 
basic properties of the surveys used for this purpose. Results are given in Section 3, divided into stars, galaxies and quasars, 
while the conclusions are given in Section 4.   

\section{Confronting Alerts with Archive Data}

 We have selected all Alerts appearing in the Gaia Alerts website %(http://gsaweb.ast.cam.ac.uk/alerts) 
 from the beginning (first one Gaia14aaa, detected on August 30th, 2014), to the end of October 2018 
 (last one Gaia18dge, detected on October 31st, 2018), that is a total  of 6308 candidates.  
 \cite{kostrz18} have investigated the detectability of nuclear transients by Gaia
  using a method different from the one used for the standard Alerts (AlertPipe) and found 482 candidates in 
 the period ranging from June 2016 to June 2017, only 5 of which were also detected by the standard Gaia AlertPipe  system.  
 We  have included in our search all candidates from this GNT list. As these authors have, by definition,  targeted galaxies only, 
by cross-matching on source position a Gaia source with the sample of SDSS-DR12 \citep{Alam2015}
  catalogued galaxies or quasars,   each object from their sample already has one associated SDSS entry, but not necessarily a spectrum for classification (they have only 142 spectral classifications, out of 482 objects). We therefore extended the search to find possibly other spectra and characteristics of their objects,  and look for long term variability.  
  
Our sample has been cross-correlated with the LAMOST DR5 database\footnote{http://dr5.lamost.org}, see also \cite{Luo2015}, 
 using a search radius of 3$\arcsec$. LAMOST is a quasi-meridian reflecting Schmidt telescope, with  an effective aperture 
between 3.6 m and  4.9 m (depending on the declination and hour angle of the pointing) and a field of view of 5 degrees  diameter 
\citep{Cui2012, Wang1996, Su2014}.  
With a wavelength coverage from 3700 to 9000$\AA$ and a spectral resolution of R $\approx$ 1800, LAMOST can  observe
up to 4000 objects simultaneously. In July 2017, LAMOST has finished its first five-year regular survey, collected 
about 9 millions of spectra  released in DR5, including $\sim$8 millions of stars, $\sim$150 thousands of galaxies,
$\sim$50 thousands of QSOs and $\sim$640 thousands of unclassified objects. The second five-years regular survey started 
in September 2017. 

The flux calibration of the LAMOST spectroscopic survey is relative only \citep{Luo2015}. Comparing a sample of  targets in common with the 
SDSS indicates that the accuracy of the LAMOST flux calibration is about 10\% at wavelengths between 4100 and 8900\AA, but decreases
to 15\% at both ends due to the rapid decline of the instrument throughput \citep{Du2016, Xiang2015}.     
 The spectral response curves for individual LAMOST spectrographs can differ by up to 30\% for a given night, and sometimes more from night to 
 night  \citep{Xiang2015} but are always calibrated. The final spectral distribution may also show some minor bumps between 5700 and 5900\AA, in 
 the connecting region between the blue and the red spectra.   
 So when comparing  multi-epoch spectra, these uncertainties at the blue end and in the 
overlap region should be taken into account. 

The limiting magnitude for the Galactic survey was r$\sim$17.8, going down to 18.5 mag in some limited fields, see \cite{Yuan2015} for more details. 
For the Extragalactic survey, the limiting magnitude for galaxies (that is, extended objects) was approximately r $\sim$18, see \cite{Shen2016} for example; 
 for point like objects like QSOs,  
the limiting magnitude  was $i=20$, see \cite{Ai2016} and \cite{Huo2015} for more details. 
%({\bf Still need here to describe the characterisitics of the spectra, including the break around 5900 $\AA$ , and the limiting magnitude of 18th}) 
Since the LAMOST field of view is circular, field overlapping was necessary to achieve a continuous sky coverage. 
During the survey, some regions of the sky were therefore covered several times, and, as a consequence, some of the Gaia objects have 
been measured several times also (3 times on average). Although the ground-based observation is 
rarely coincident with the alerting date, this opens the possibility to look for intrinsic variability independently 
 of the Alert from Gaia. 

In order to complement this, we have thus also cross-correlated our sample with the Sloan Digital Sky Survey 
 (SDSS) sample \citep{Aguado2019, york00}, up to DR15,  to provide more observations, using the same search radius,   
 and this yielded a larger number of  identifications.  
  The DR15 is the latest Data Release from SDSS-III \citep{Aguado2019} and the SDSS globally provides spectra with 
  an average resolving power of R $\sim$ 2000, from $\sim$ 1300 at the blue end (3600$\AA$) to $\sim$ 2500 at the red 
  end ($10000\AA$) \citep{smee13}.  The spectra are divided in two parts, blue and red, 
 with a dichroic  splitter at 6000 $\AA$, giving thus a small uncertainty around this wavelength. The entrance fiber sampled 3$\arcsec$ on the sky at the beginning of the SDSS survey
  \citep{york00}, from DR1  to DR8, and then switched to a 2$\arcsec$ entrance fiber with the start of the 
  BOSS survey \citep{dawson13}, from DR9 to the present release. 
  The search radius of 3$\arcsec$ used to find counterparts to our Gaia Alerts is thus consistent  with the ground-based surveys  properties.  
   
  Spectral information for the Gaia Alerts and the GNT sample are given in the Tables. A few lines are given as example   
   in the Appendix (with description of the different columns),  
   while the full Tables are available online only.
   
\section{Results}

\subsection {Stars and Cataclysmic Variables} 

From the Gaia Alerts studied here, sixteen have single epoch SDSS or LAMOST spectra labelled as ``STAR". 
 There are none among the GNT list, obviously because those alerts targeted, by definition, only galaxies. 
 Among  these sixteen, there are two late-type stars: Gaia14aaq, an M-type star with a SDSS spectrum, was also classified by \cite{jonk15} 
 during one of the first ground-based follow-ups of Gaia Alerts; 
 and Gaia17bsx, an M-type carbon star with a LAMOST spectrum, but already classified in an earlier survey by \cite{maur04}.  
 Two others look like early type stars, as indicated from their LAMOST spectra: Gaia14adz has a spectrum labelled as G6 star, and its Gaia light curve shows no concluding features; 
 and Gaia17akp was labelled A1V in  the LAMOST DR5 Version 1 release, but its spectrum, while showing mainly hydrogen lines and Na in absorption, displays also  a small H$\alpha$ emission   and further faint Balmer emissions are seen inside the broader, higher order, Balmer absorption lines, and is therefore possibly a CV (revised as such in the DR5 Version 3 release). 
 Two others are white dwarfs from their SDSS spectra, also confirmed in Simbad:  
 Gaia14aab is a sdA sub-dwarf from \cite{Kepler2016}, while Gaia14adk was confirmed as a white dwarf by \cite{Eisenstein2006}.
 But as none of those spectra are contemporary to the Gaia Alerts, it is difficult to assess what caused the alert, 
 and how it would have reflected in their spectrum. 
 Two others are mis-labelled stars: Gaia16acx, alerted as a SN candidate by Gaia on Feb. 9th, 2016, has a LAMOST spectrum 
 within a distance of 2.75$\arcsec$, taken on Dec. 25th, 2012, which was mis-labelled as  "STAR" due to its low SNR, 
 but corresponds in fact  to the  galaxy UGC 691 at the redshift of 0.04907  \citep{Mahdavi2004}; 
 it is thus not contradictory with the Gaia Alert being indeed due to a SN in this host galaxy; 
 and Gaia17bqd,  labelled as  ``STAR" in SDSS, is in fact a BL Lac object at z=0.7269 identified by \cite{Paris2014}, quoted by Gaia Alerts, but the spectrum 
 is very flat and the emission lines are barely seen, the redshift is highly uncertain. 
 Here also, the large time interval between the Gaia Alert (June 25th, 2017) and the date of the SDSS spectrum (Jan. 2nd, 2003) 
 does not allow to conclude on the nature of the observed variation.   
 
\begin{figure*}
%\centering
  \includegraphics[width=3.4in]{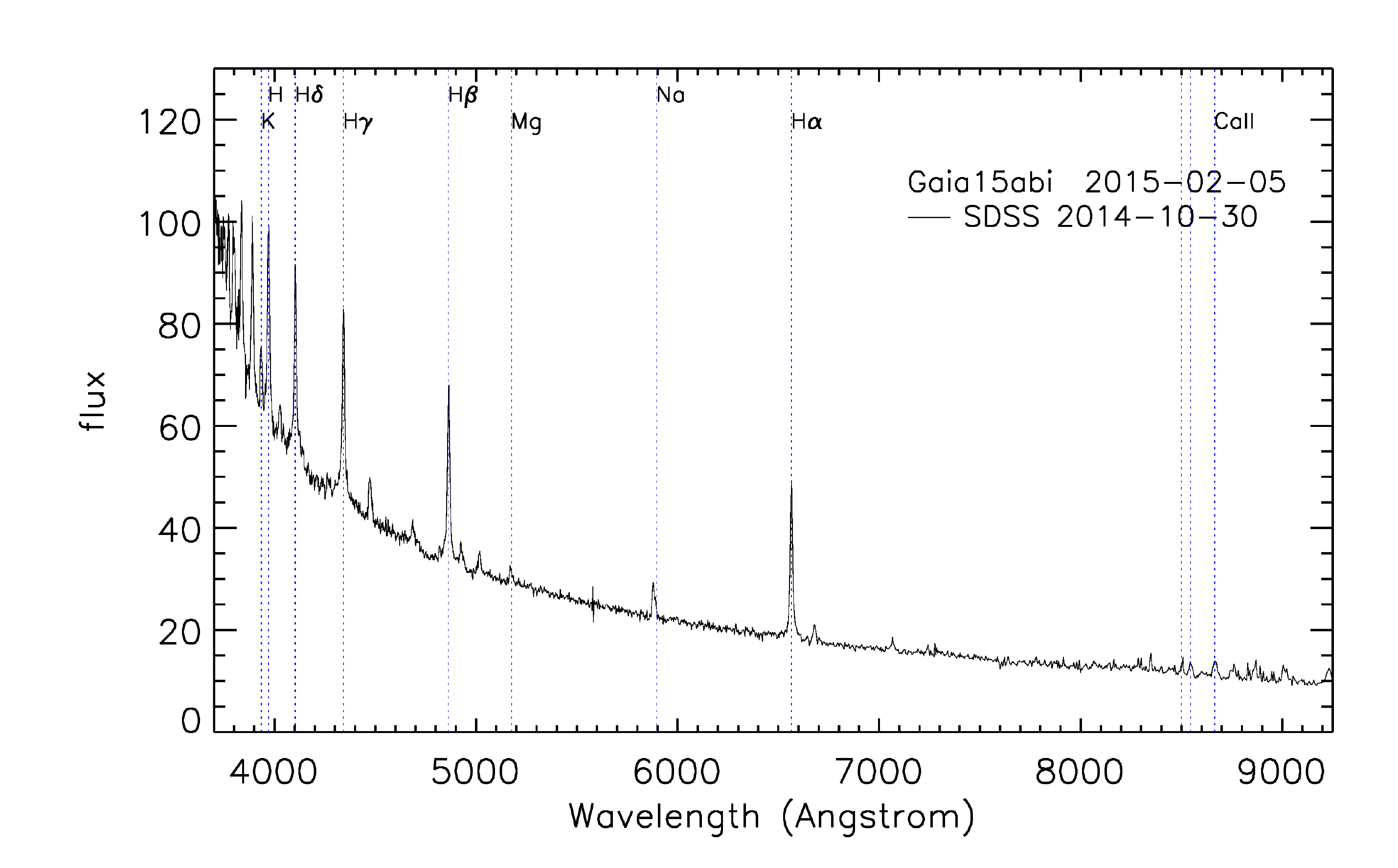}   % new CV, single-epoch
  \includegraphics[width=3.4in]{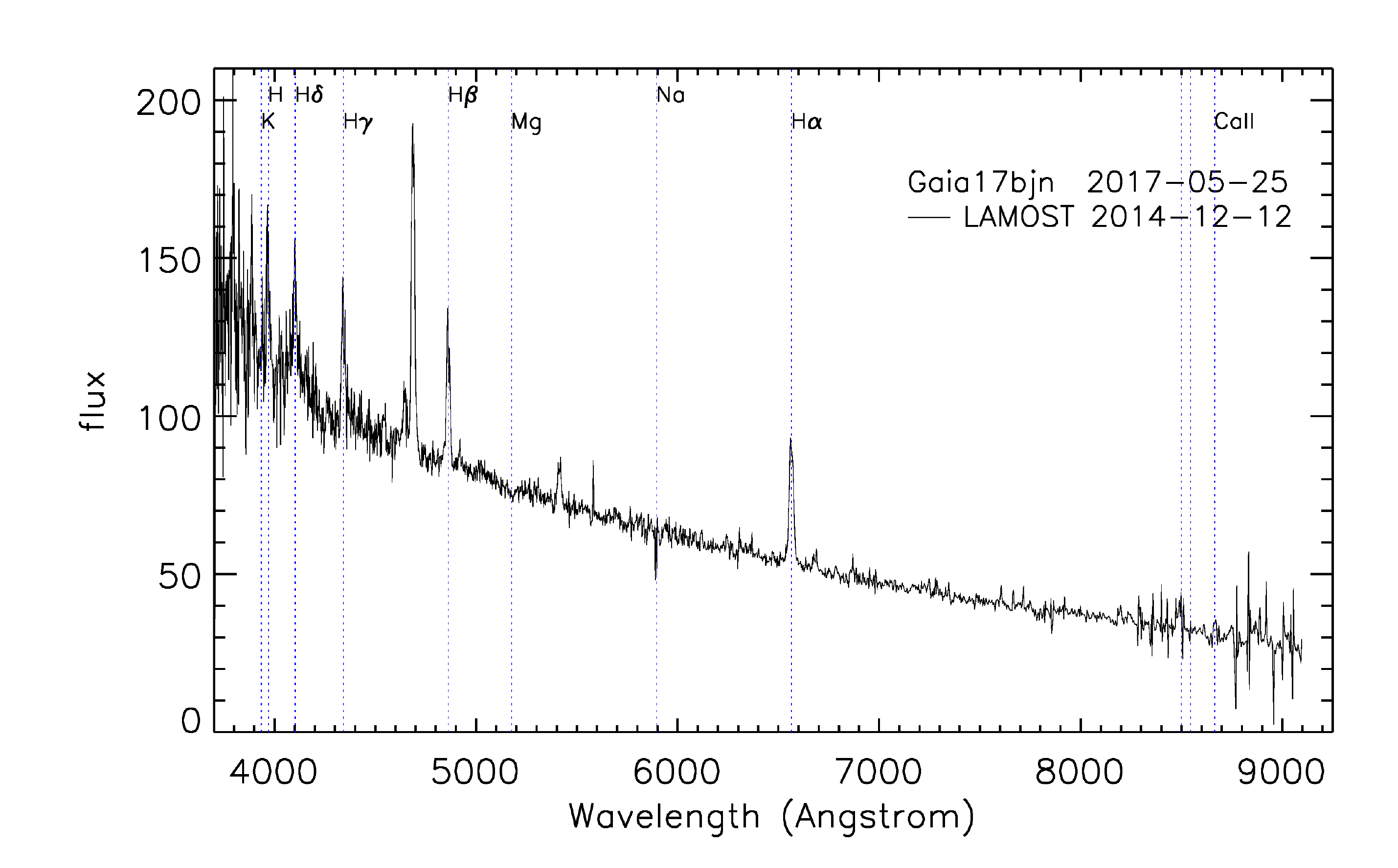}   % new CV, single-epoch
  \includegraphics[width=3.4in]{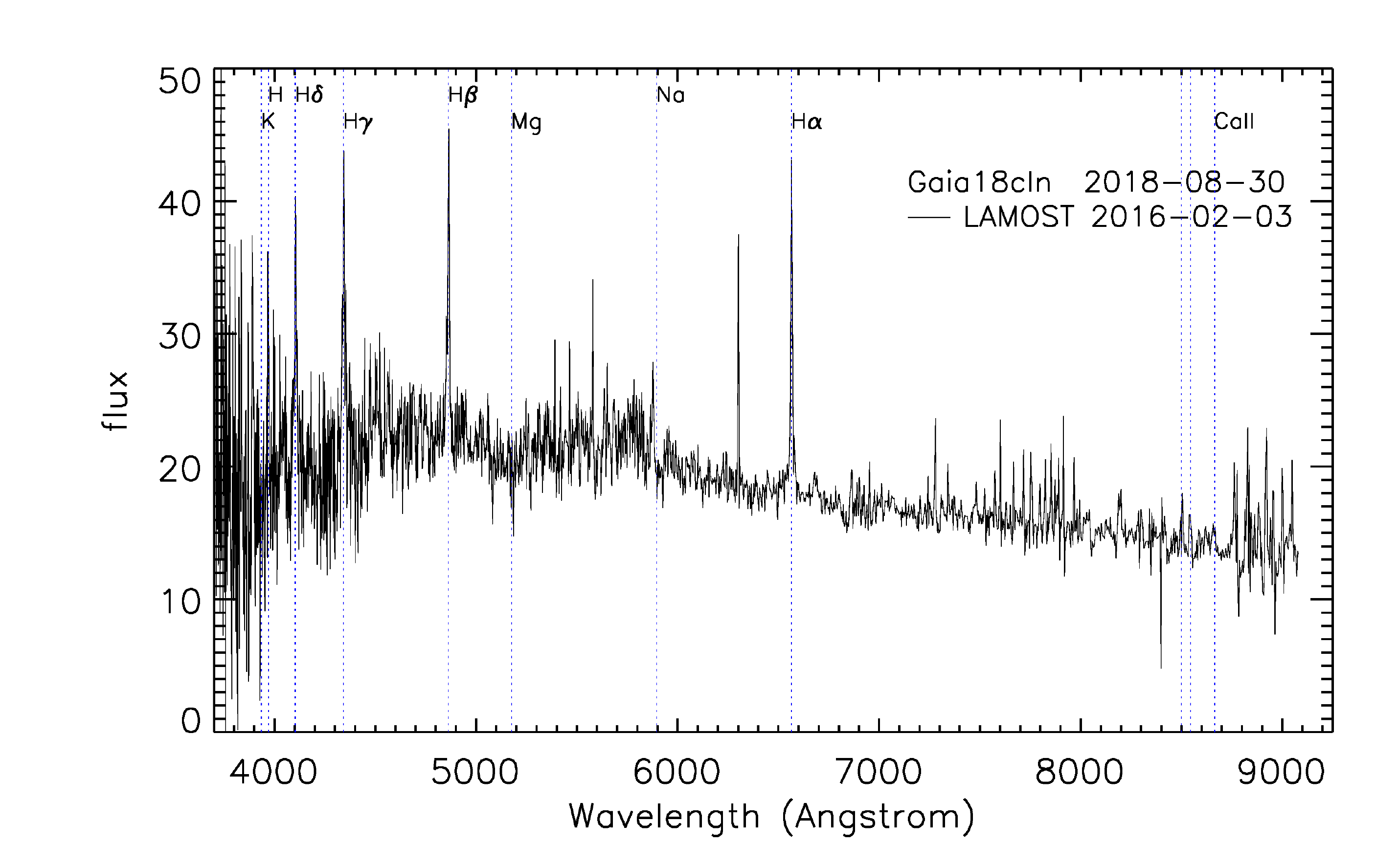}   % new CV, single-epoch
   \includegraphics[width=3.4in]{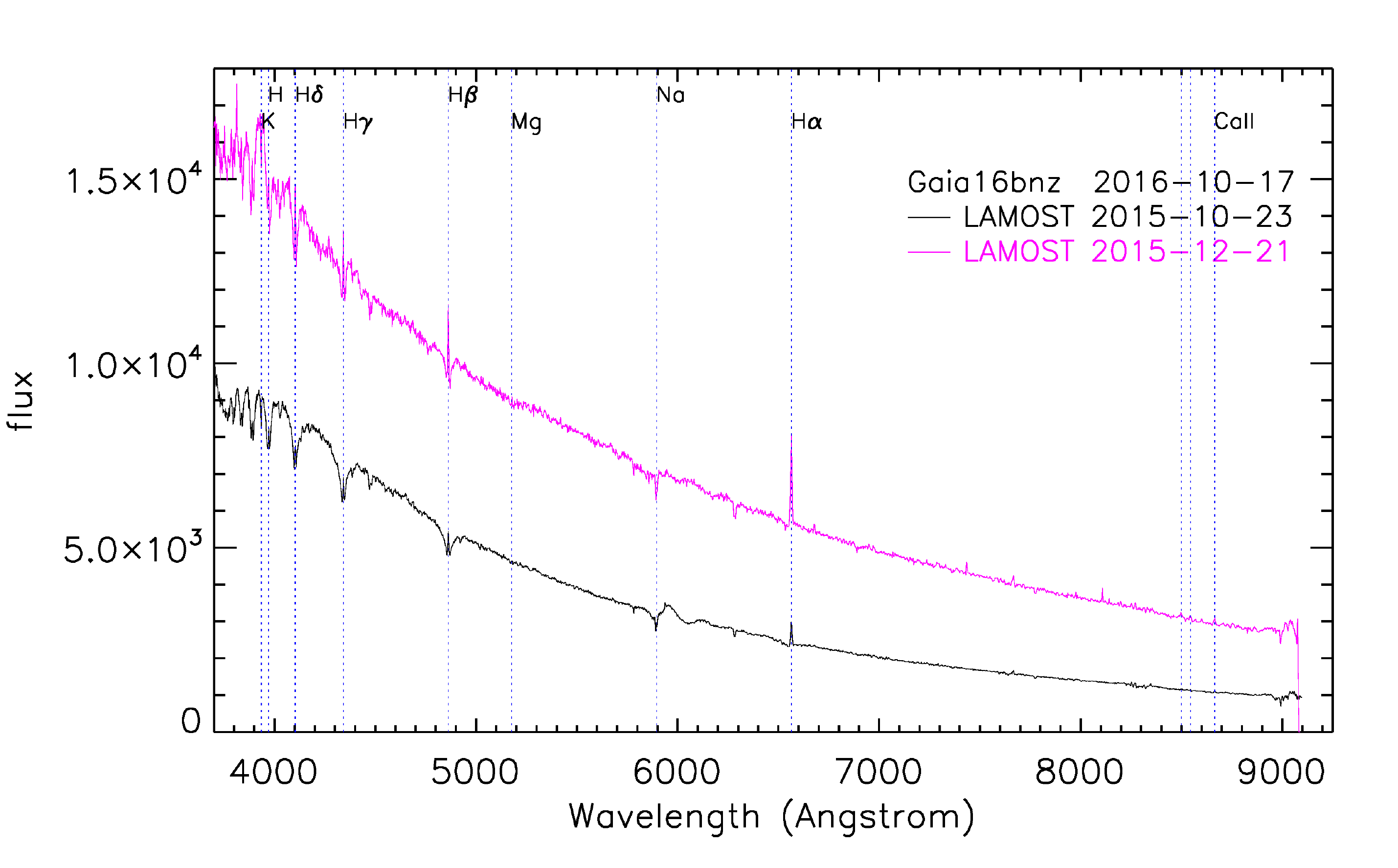}  % CV, multi-epoch
   \includegraphics[width=3.4in]{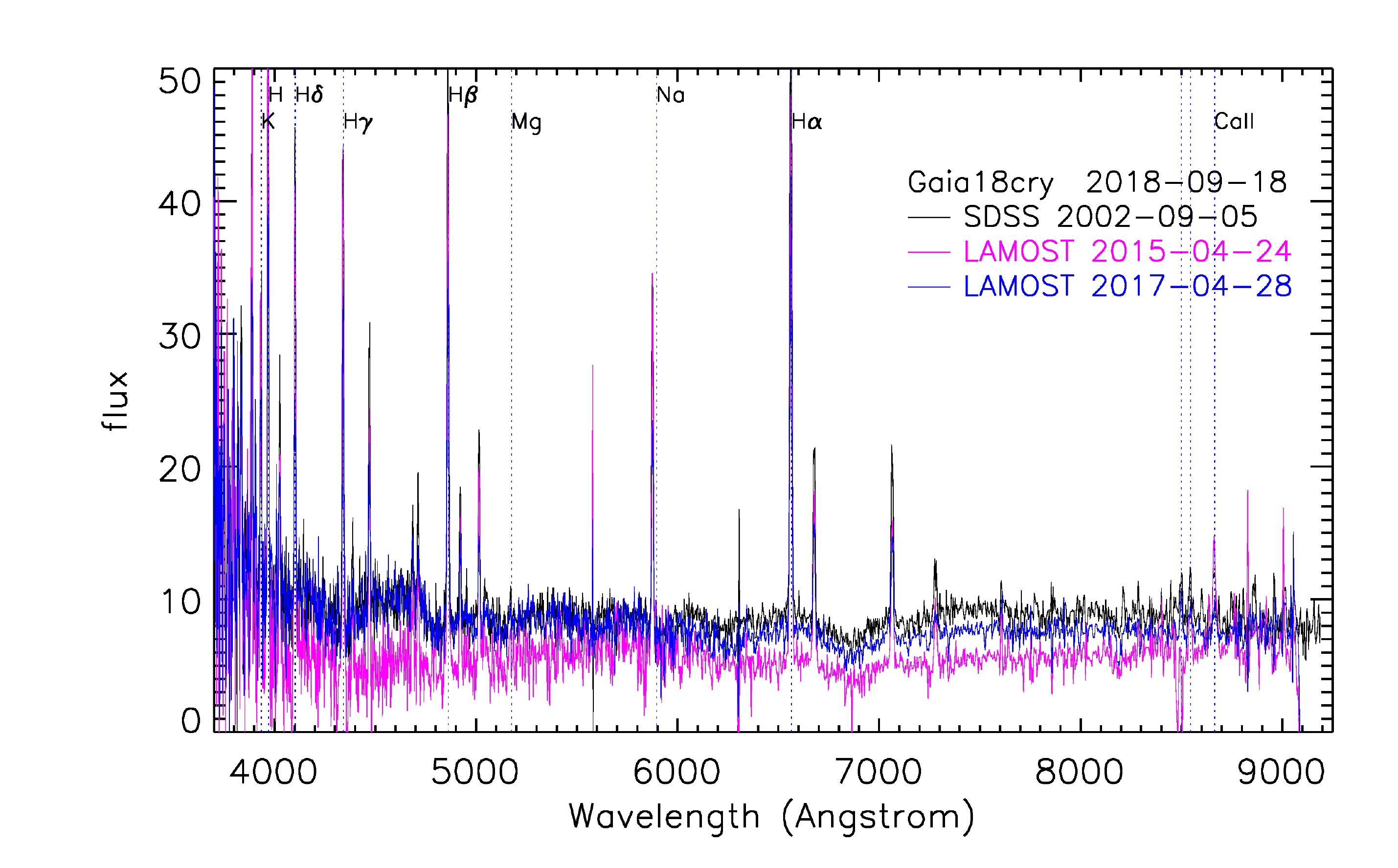}  % CV, multi-epoch
  \includegraphics[width=3.4in]{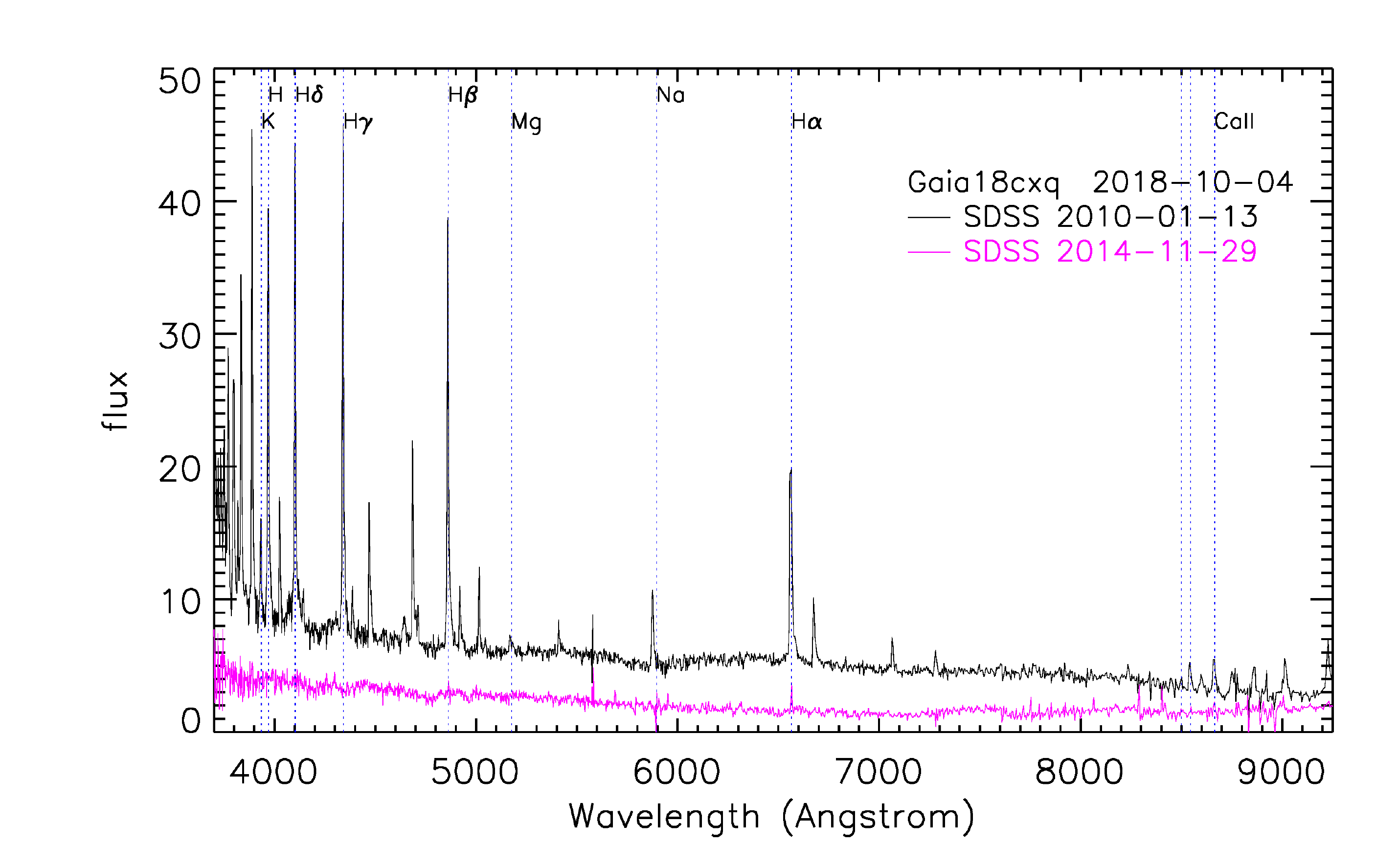}  % CV, multi-epoch
  \caption{\label{fig:star_CV} Some representative spectra of CVs from the Gaia Alerts. From left to right and top to bottom: 
  Gaia15abi, Gaia17bjn, Gaia18cin, Gaia16bnz, Gaia18cry and Gaia18cxq. The dates of the Gaia Alert are given in each figure, 
  together with the dates when the spectra were obtained. Typical lines are labelled. }
\end{figure*}

 All the eight others are Cataclysmic Variables (CVs). Five of them were already known, and include:  
 Gaia16apa = CRTS CSS090918 J001538+263657, a known CV from the Catalina Real-time Transient Survey 
 (CRTS; \citealt{Thorstensen2016, Drake2014a, Szkody2014}); 
 Gaia18bvj = V521 Peg, a variable star in \cite{Kholopov1998};
 Gaia18bwz = QZ Vir (also named T Leo, \citealt{Kato1997}),  a known dwarf nova in outburst caught by Gaia in July 2018;  
 Gaia18crs = QW Ser, a known SU UMa-type dwarf nova with a 5 magnitude  brightening outburst  caught by Gaia  
 \citep[see][]{Patterson2003, Nogami2004, Szkody2009};
 Gaia18cwe = V493 Ser, a known CV from the CRTS \citep{Drake2014a}. \\
  Three CVs are newly identified: Gaia15abi, Gaia17bjn, and Gaia18cln, their SDSS or LAMOST spectra are shown in Figure \ref{fig:star_CV}. 
  
  All of those CVs, except one, show the classical Hydrogen and Helium lines in emission, 
 with the addition sometimes of the $\lambda\lambda\AA$ 5169 FeII, 4645 NIII-CIII or 4267 CII lines. 
 The exception is the known CV QWSer = Gaia18crs 
 which displays only strong Balmer absorption lines, but the SDSS spectrum is from 2006, thus many years before the Gaia Alert of 2018.  The spectrum would thus indicate that the object was  caught in a high-accretion phase at that time, while no spectrum is available during the Gaia Alert. Following this line, 
 it is probable that the stellar spectrum mentioned earlier for Gaia17akp, with Balmer absorptions and only  
 weak Balmer emissions, taken by LAMOST about two years before the Gaia Alert, indicates in fact also a 
 CV but close to an outburst state. The  points of the Gaia light curve start only on March 29, 2015, that is about 2 months after the LAMOST spectrum was taken, and remain low until the first alert on 13th Feb. 2017, but several outbursts are then seen over the next two years, confirming its CV nature. 

In addition to those single-epoch spectra, thirteen objects alerted by Gaia have multi-epoch spectra labelled as ``STAR". Among them: 
Gaia14acd is an F type star classified by LAMOST, the reason for the alert is still unknown (was not reobserved since then).
Gaia14ado is a WD+MD binary with 2 epochs SDSS spectra \citep{Eisenstein2006}.
Gaia16bft is a young stellar object in the Simbad database.
Gaia18asf is an M type star with emission lines seen in the  SDSS spectra.
And Gaia18bla, which has a SDSS spectrum labelled as star, is in fact a BL-Lac object at redshift z = 0.212 \citep{alb07}, correctly identified by Gaia Alert due to the coincidence in position.  
 
 Eight objects are CVs, which include Gaia16adh, 16ahb, 16ahl, 16bnz, 18cqo, 18crc, 18cry and 18cxq.
 Gaia16adh is a CV with an M type donor, confirmed by \cite{Kepler2016} who used the SDSS spectra, which were taken in 2013 and 2015. 
 This object had already been  pointed out as a SU UMa-type candidate from its photometric behavior 
 \footnote{ http://ooruri.kusastro.kyoto-u.ac.jp/mailarchive/vsnet-alert/1231}  but is not in the SIMBAD database.  
 Gaia16ahb is a CRTS variable identified as CV candidate by \citeauthor{Drake2014a} (2014a, 2014b) and spectroscopically confirmed. Its light curve shows only one eruption on March 1st, 2016 during the whole Gaia survey until end of 2019. 
 Gaia16ahl = IW And is a Nova-like star known from photometry, and confirmed as CV by \cite{Gentile2015} using LAMOST spectra: 
 it was observed by LAMOST from 2012 to 2017  four times, and Balmer absorption/emission lines in different states were  detected; 
 it is the Z Cam system IW And as noted by Gaia Alert. 
 Gaia16bnz is a new CV with 2 epochs LAMOST spectra taken in 2015 October and December: Balmer absorption/emission lines
 are present and changed between the two dates, which are however anterior to the Gaia Alert of Oct. 17th, 2016 (Figure \ref{fig:star_CV}).  Its Gaia light-curve does not display strong variations during the $\sim$ 5 years of survey, the Alert being issued because of a long-term decline, and was thus not readily identified as a CV.  While the first LAMOST spectrum shows essentially broad Balmer absorption lines, the second spectrum about 2 months later shows a strengthening of the Balmer emission lines within weakening absorptions, consistent with a progressive return to the low state. 
 Gaia18cqo = TX Tri is a CV identified by  CRTS (\citeauthor{Drake2014a}, 2014a).  
 Gaia18crc =TW Tri is a CV with spectra taken at 5 different epochs  by LAMOST, none coincident with the Gaia Alert, 
 the object was investigated by \cite{Thorstensen1998} and \cite{Gentile2015}.  
 Gaia18cry is a CV (catalogued in SIMBAD), with spectra at 3 different  epochs, one taken by the  SDSS and two by LAMOST.  
 And finally Gaia18cxq = EG Lyn is a CV (catalogued by SIMBAD) with 2 epochs SDSS spectra, see Figure \ref{fig:star_CV}.
 For none of those are the LAMOST or SDSS spectra contemporary to the Gaia Alert. 
 % refer to outbursts in Cataclysmic Variables: 
%two of them were known (G18cqo = TX Tri and G18crc = TW Tri), 
%while the three other ones are confirmed as CV by their spectra. 

The spectrum of Gaia16adh is remarkable as the Balmer emission lines are double-peaked: the Full Width at Zero Intensity is 
measured at 4300 km/s for H$\alpha$  (or 4380 at H$\beta$), and the FWHM is 2200 km/s. This is quite typical of accretion disks viewed at high inclination and confirm the CV nature.  
There is no significative difference seen between the two SDSS spectra of, respectively, 
18th January 2013 and December, 1st, 2015, but we have no data to estimate whether the object remained in a high state during this interval.  The Gaia outburst detected later on February 11th, 2016 
was a short, but bright outburst, but no further spectrum is available at, or after, this event.  The next outburst seen in the light curve is only on Feb.5th, 2020, that is about 4 years later (compared to 3 years between the two spectra): it is thus quite plausible that the period of recurrence of this object is about 3-4 years. 
A few other spectra of CVs (like Gaia16ahb, 18bvj or 18cwe) mentioned above show a similar, 
double-peaked emission line profile, the most notable of them being V521Peg = Gaia18bvj, whose SDSS spectrum shows a FWHM of about 2000 km/s, but was taken about 4 years before the Gaia Alert of July 2018, no other outburst being recorded during the period covered by Gaia. 
It would obviously be quite desirable to get regularly spaced spectra of those objects  to monitor their evolution. 
 Discarding the miss-identified spectra of galaxies or quasars, we can conclude from this sample that a large fraction 
of the Gaia Alerts where the corresponding SDSS or LAMOST spectra were labelled as stellar are in fact CVs. 

\subsection{Galaxies} 

Among the Gaia Alerts, hundred and fifty-seven have been found associated with a galaxy where a single epoch 
 spectrum was available in either LAMOST or SDSS. From the GNT list of 
 \cite{kostrz18}, forty one are in the same situation, giving us a total of 198 galaxies with single epoch spectra
 \footnote {Note that although the initial GNT sample comprises 482 targets selected as having a SDSS counterpart, 
 only about one third of them have a SDSS spectrum for classification, the rest of the sample having only SDSS photometry.}.
 Among those galaxies, excluding 4 where the spectra are of very low SNR \footnote{These are Gaia17aqc, a blue source with previous variability in CSS, too faint for LAMOST, with a photometric redshift of 0.925; Gaia17bmd, a UV and radio source with no measured redshift, but a single epoch LAMOST spectrum shows a single emission line giving z = 0.159 if identified to H$\alpha$, to be confirmed; Gaia17cdu,  a candidate SN in a LEDA galaxy with cz = 31028 km/s, alerted at g =18.9, giving an absolute magnitude of -19.3, compatible with a type Ia SN; and GNT J000121-0011, a NED galaxy with z = 0.462, but no identifiable features in the SDSS spectrum.},  
only 22 (Gaia Alerts) + 7 (GNT list) 
  show no emission lines in their spectrum, the remaining 132+33 galaxies all having  emission lines of various types and strengths.  
  For most of the galaxies, some star formation is therefore still present, consistent with the idea that 
  the Alert was often caused by the explosion of a SN. 
  However, some of them show a broad H$\alpha$ component underlying the narrow emission line, 
  while no broad component is detected under H$\beta$ (examples are Gaia14adg, GNTJ115000+3503, or GNTJ140030+5653, under others). 
  The broad H$\alpha$ component is strong in only a few of them,   
 but even then no broad H$\beta$ component is apparent. These objects are therefore all Seyfert galaxies of type 1.9. 
 In those cases, the variability detected by Gaia could well be due to changes in the accretion rate or in the obscuration in front of 
 the AGNs. In favor of this interpretation is the fact that we have proportionally 
  more cases of broad lines in the sample of galaxies detected in the GNT list than in the standard 
  Gaia Alerts ($\sim$ 27\% versus $3\%$). 
  The fact that only Sey 1.9 are seen here could lead to the conclusion that the variations are due to changes in obscuration only: 
  this is however misleading and probably a selection effect.  
  Indeed, variations in Sey 1 or quasars are seen also (see next section), and the above objects appear only in the list of   ``Galaxy" because their Seyfert properties were not recognized in the surveys. \\

\begin{figure}
   %\centering
    \includegraphics[width=3.1in, height=2.2in]{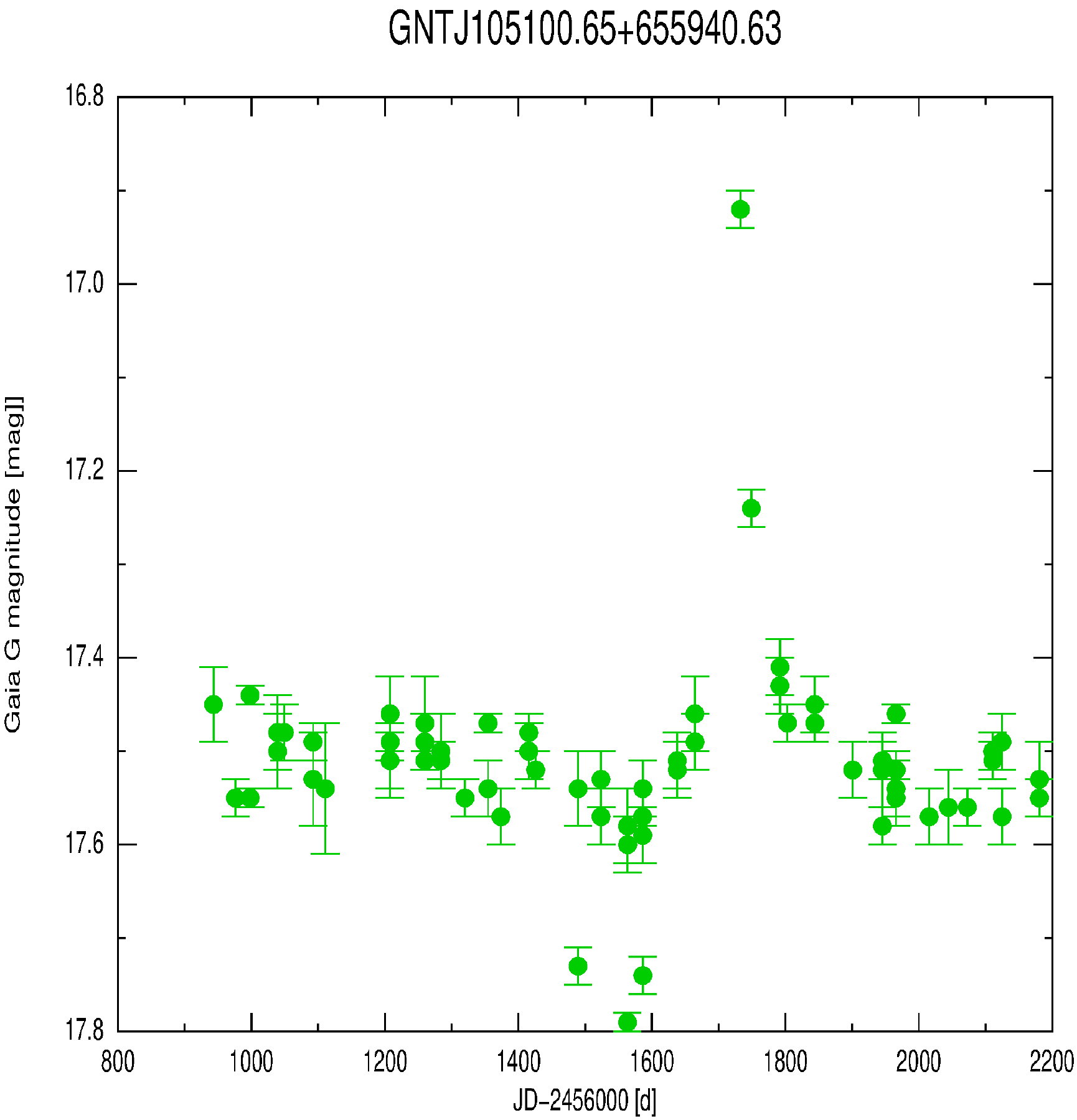}
    \caption{The light curve of the galaxy GNTJ105100+6559, as given by \cite{kostrz18}, looking like a SN outburst. 
    \label{fig:GNT1051_LightCurve}}
   \end{figure}

\begin{figure}
   %\centering
    \includegraphics[width=3.3in]{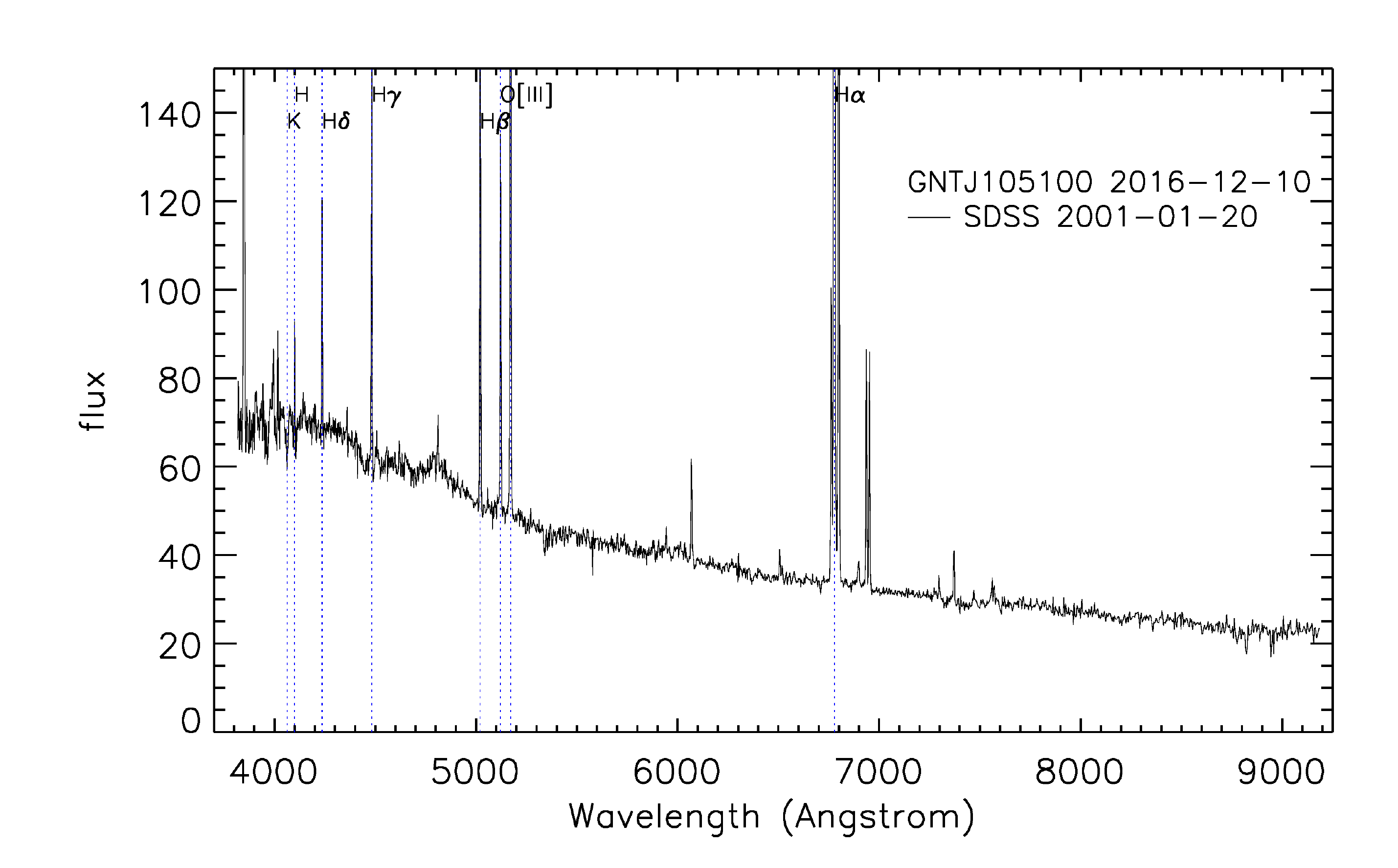}
    \caption{The galaxy GNTJ105100+6559, showing strong emission lines and a blue complex around $ \lambda\lambda  4650 \AA$, 
    characteristic of the presence of WR stars. This is a star-forming galaxy. 
    \label{fig:GNTJ105100}}
   \end{figure}

 We have therefore also looked if some of the emission line objects could  be AGNs, particularly type 2 Seyfert galaxies. Type 2 Seyfert galaxies are easily detectable because they have a strong $[\rm{OIII}] 5007\AA$ line compared to H$\beta$, and a comparatively strong 
 $[\rm{NII}] 6584\AA$ with respect to H$\alpha$, ratios which are quasi independent of the reddening. To quantify this, 
we used the well known BPT diagram \citep{BPT1981} in its revised form from \cite{Kewley2006}, derived from a large sample of SDSS spectra. This allows also to detect ambiguous cases, Liners or composites (a mix of Star Formation and AGN). At the end, very few emission line objects clearly appear in the Seyfert part of the diagram: e.g. Gaia17bip , and  GNT014153+0101  (both classified as AGN broad line by SDSS, but there is no clear broad line component in their spectra! so they are rather Sey 2), or Gaia14aak, Gaia18crv,  GNT080115+1101, GNT081437+1722 and GNT120346+5100 which are all genuine Seyferts. A  few others fall at the limit between Starbursts and AGNs or Liners, like Gaia18dbt, or GNT131839+4630,   
or have an ambiguous classification because, although their [NII] is strong,  the H$\beta$ line is not well detected above the continuum, such as Gaia14aaa, Gaia18cuj or GNT003719+2613.  Interestingly, in a few cases, although the galaxy was classified as active, the Alert was nevertheless probably due to a SN, as seen from the light curve (e.g. Gaia17bip, or Gaia18crv (spectroscopically classified SN2018gho), or GNT081437+1722).  For the other AGNs, the light curve is more complex and points towards intrinsic variations. \\
For all the other emission line objects, the starburst nature is therefore clear, and in most cases the Gaia light curve is compatible with a SN.  
 We highlight one example, GNTJ105100+6559, which shows  a strong emission-lines spectrum typical of starburst galaxies, but  also  
  a ``blue bump" around $ \lambda\lambda 4650\AA$ characteristic of a Wolf-Rayet (WR) contribution. This galaxy is thus forming massive stars and  
 it is then plausible that  Gaia has detected a SN, even if the light curve does not allow to establish its type. The Gaia light curve data (although poorly sampled) are given in \cite{kostrz18} and shown here in Fig. \ref{fig:GNT1051_LightCurve}, together with  its spectrum in Fig. \ref{fig:GNTJ105100}.

  But to be able to really identify the source of variation in all those galaxies, spectra close to 
  the detected Gaia variation would be essential, together with a reference spectrum at a different epoch.  
  We have therefore looked with particular attention to those galaxies where several spectra were available in the 
  databases. We have found 63 of those, including 13 from the GNT list. 
  Gaia14aak and Gaia14abw are distant by only 0.58$\arcsec$, and  correspond  to the same target in
  the SDSS and LAMOST archives, a type 2 Seyfert AGN with z=0.064\footnote{They probably also correspond to a unique Gaia target, but detected at different scanning angles, not recognized as such in the very early phase of the survey.}. So the final sample of galaxies with multi epoch spectra
  contains  62 objects.  
  
\begin{figure*}
 %  \centering
   \includegraphics[width=3.4in]{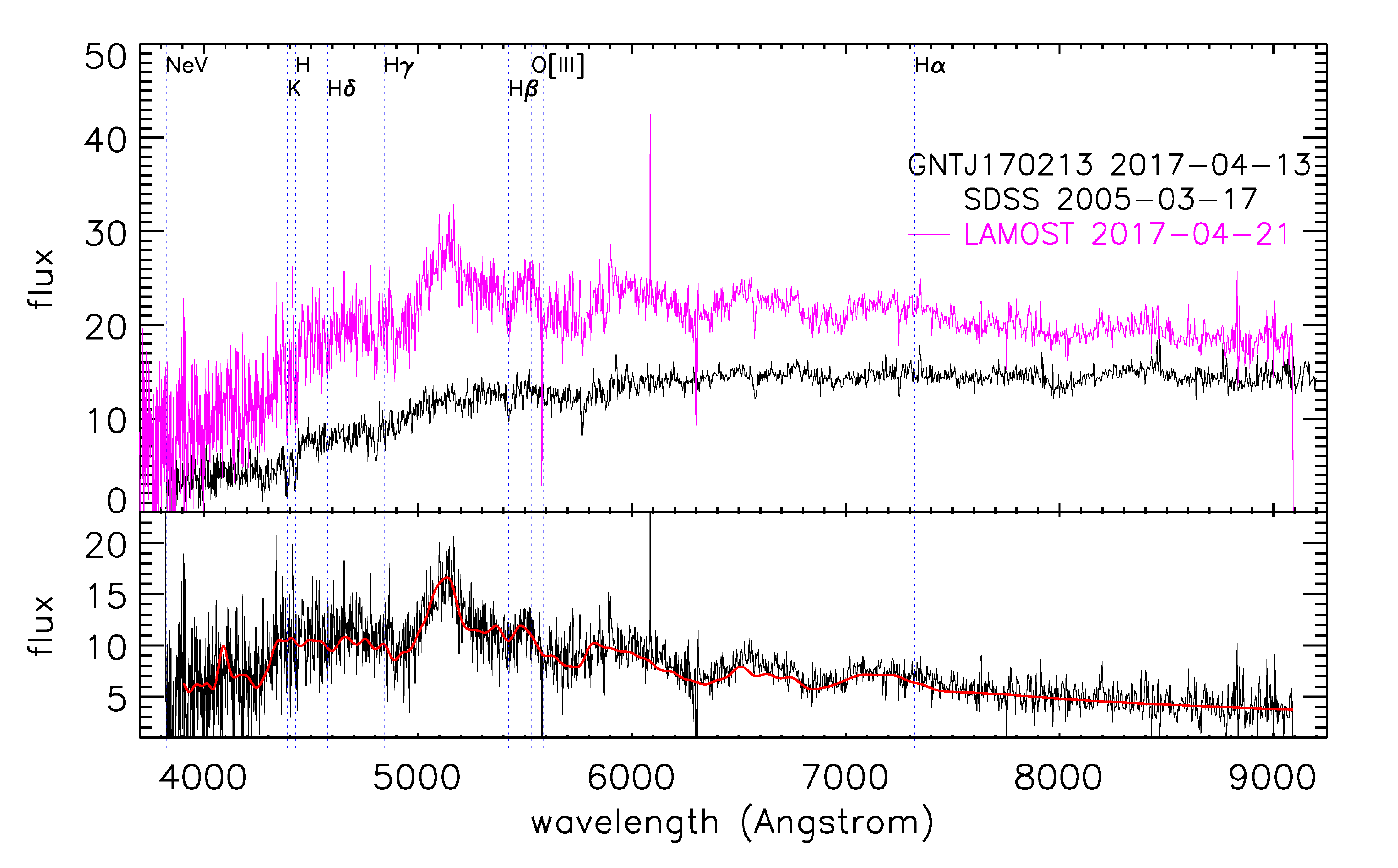}
   \includegraphics[width=3.4in]{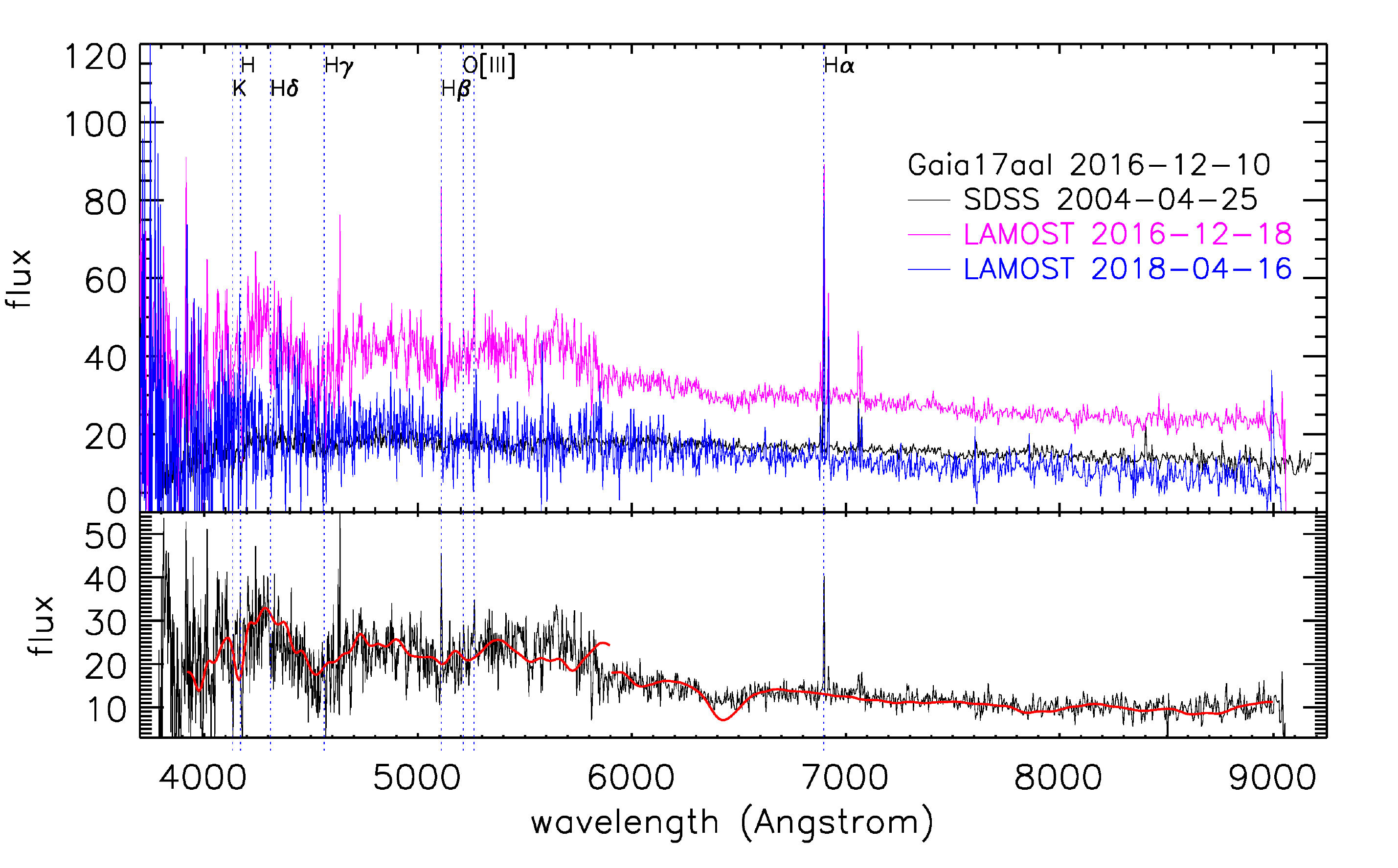}
      \caption{Left: Spectra of GNTJ170213+2543 observed by SDSS on March 17th, 2005
      and by LAMOST  on April 21st, 2017 (top panel); residual after subtraction of the earliest spectrum from the later one, 
      showing a type Ia SN spectrum,  overlaid with the fit from SNID (bottom panel).
      Right: Spectra of Gaia17aal observed by SDSS on Apr. 25th, 2004 and by LAMOST on Dec. 18th, 2016 and Apr. 16th, 2018 (top panel); 
      residual after subtraction of the earlier SDSS spectrum from the LAMOST spectrum taken in 2016 Dec, 
      overlaid with the fit from SNID (bottom panel, also a type Ia SN).
      }
    \label{fig:galaxy-SN}
   \end{figure*}

  However, in practically all those cases, the available spectra were taken at an epoch far away from the 
  date of the Alert, therefore bringing no clues on the cause of Alert. 
  There is therefore no surprise that we found no evidence of a SN residual in their spectrum, 
  except for two cases.  
  One is GNTJ170213+2543, where the LAMOST spectrum was obtained on April 21st, 2017, 
  that is one week after the increase was noticed by Gaia (from G=19.5 to G=18.6). We subtracted the reference SDSS spectrum of 
  March 2005 from the LAMOST spectrum to obtain the residual presented in Figure \ref{fig:galaxy-SN}. Fitting the residual with SNID 
  \citep{snid07} shows that it corresponds to a type Ia SN 
 about 15 days after its brightness maximum. The redshift derived from the residual spectrum is z = 0.117, perfectly compatible with
 the one derived from the galaxy (z =0.1155) and Gaia thus detected the increase in brightness around 1 week  
 after the maximum only, as its time coverage is very irregular. No more recent spectrum has been acquired since. 
 
 The second case is the galaxy Gaia17aal, alerted on Dec 10th, 2016. One of 
 the LAMOST observations was taken on Dec 18th, 2016, that is only one week later and the spectrum shows significant
 bumps compared to the SDSS reference spectrum taken in 2004.  The galaxy looks ``quiet" again in the second LAMOST spectrum of 2018.
 The LAMOST spectrum of 2016  displays a significant jump at 5900$\AA$, probably caused  
  by the blue/red band combination described in Section 2, so we used the blue and the red spectra independently.   The residual 
  spectrum was obtained by subtracting the SDSS spectrum from the LAMOST 2016 one, and fitted with SNID
separately for the blue and the red band. 
 Both  fittings indicate a typical, normal, type Ia SN near its maximum brightness,
 and the redshift of 0.05 obtained from the residual spectrum is compatible with the one from the galaxy , see also Figure \ref{fig:galaxy-SN}.
 
  In most of these 62 galaxies with multi-epoch spectra, we do not see any significant change in continuum or emission lines, thus suggesting 
  that the change in magnitude noticed by Gaia was indeed due to a SN, although not confirmed here 
 (but some of them were confirmed independently at different telescopes, as noted on Astronomer's Telegrams: 
 e.g. Gaia16aty on Atel9208, Gaia16avs on ATel 8935, Gaia17bka on ATel 10352, or Gaia18cgj on ATel 11938, etc...).  
In 6 out of 62 galaxies (Gaia16cav, 17bka, 17bwd, 18afc, 18awi, GNTJ085416.+2903), that is, $10\%$ of them, 
 we do not see any emission line, not even H$\alpha$, those are therefore dominated by an older stellar population. 
%\textcolor{blue}{
In the Gaia alerts fraction of them,  in only two cases (4\%), Gaia17bcr and 17dbr,  do we see a weak, broad H$\alpha$ component (but for these two, the light curve is more indicative of a SN than of nuclear variation),   
 therefore there is no reason to suspect  AGN variation as the main cause of the magnitude change in this sample. 
%}
 On the contrary,  in the GNT list, this proportion is much higher, 5 galaxies out of 13 with multi-epoch spectra ($\sim$40\%) : GNTJ084157+0526, GNTJ084535+3439,  GNTJ131101+0003, GNTJ143445+3328 and GNTJ172959 +6242  show a broad H$\alpha$ emission line. 
It is however small number statistics, further biased by the random availability of spectra in databases. But if we look 
at the  whole GNT sample, among the 146 objects (out of 482) where at least one spectrum is now available, two thirds are classified as ``QSO", as compared to only $17\%$ in the full Gaia Alerts, supporting the idea that a large fraction of the variations seen in the GNT sample are due to the central AGN itself. This is consistent with the purpose of \cite{kostrz18} who looked specifically for nuclear variations. 

There are a few cases where a change of slope is seen in the blue part of the continuum (e.g. Gaia15afx, 16ajl, 16aru, 
17bcr, 17bij or 17bwd, 17bwl, 17byx), and the increase is relatively large, of the order of at least $50\%$.
It is difficult to say whether all this change is intrinsic or partly due to losses because of atmospheric differential refraction, 
poor centering into the entrance fiber, or larger uncertainty in the LAMOST relative flux calibration, 
 specially when close to the  LAMOST magnitude limits.  

\begin{figure*}
   \includegraphics[width=3.4in]{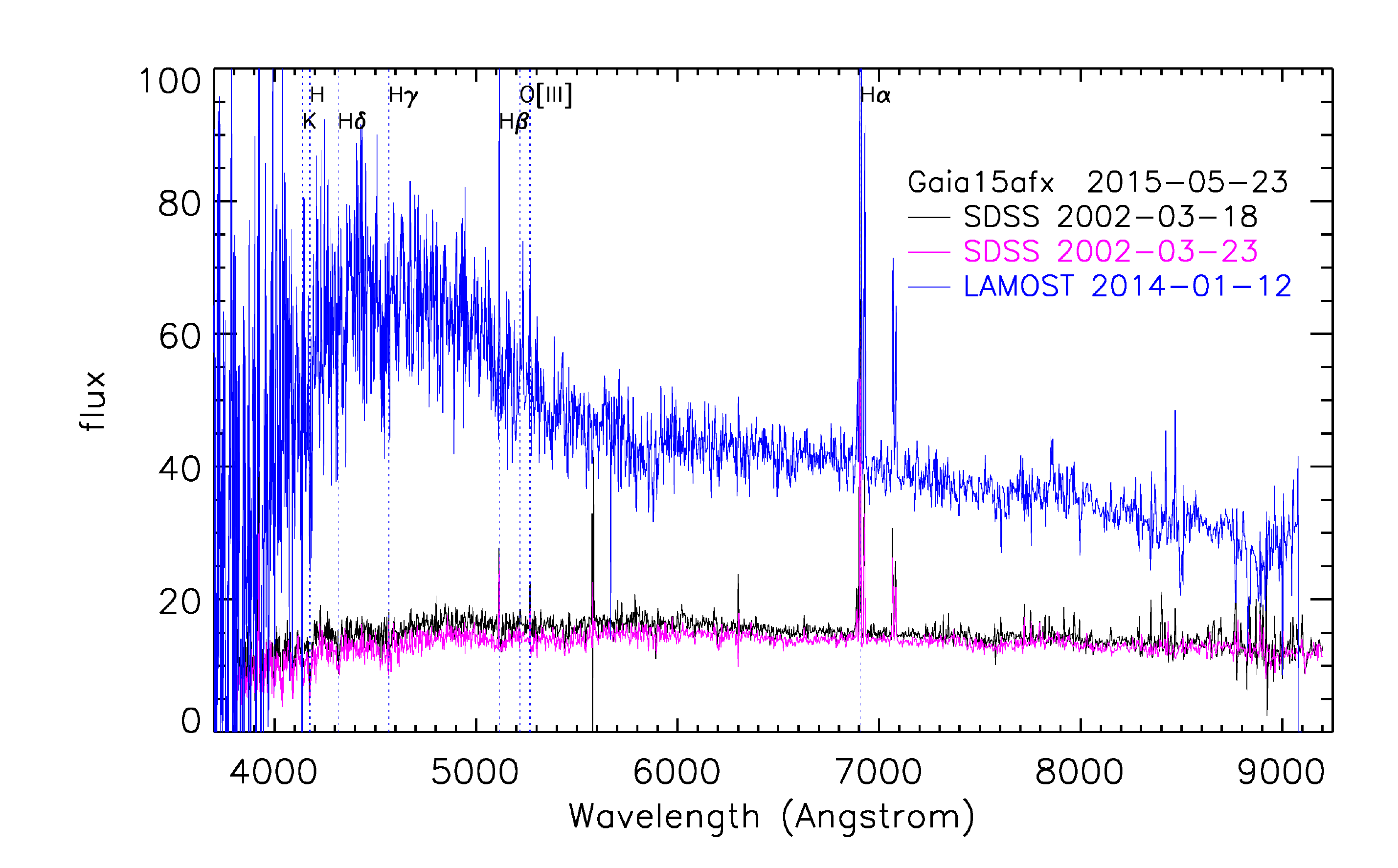}   %Gaia15afx is very nosie
   \includegraphics[width=3.4in]{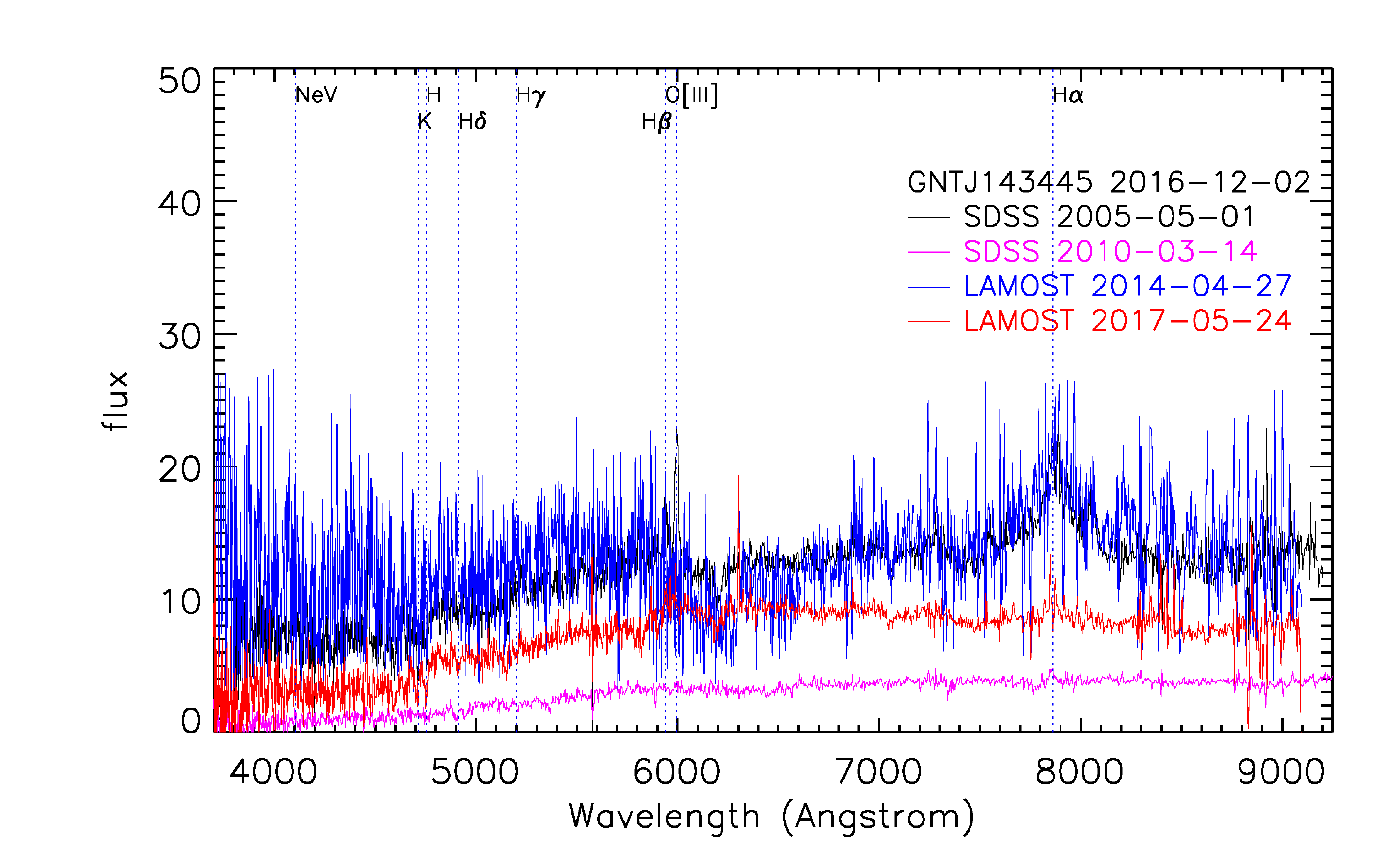}
        \caption{Left: Spectra of Gaia15afx, where the latest LAMOST spectrum (in blue) seems to indicate a change in the blue part of the continuum. 
        Right: Spectra  of GNTJ143445+3328. The lower (pink) spectrum belongs to a different object, but the latest LAMOST spectrum (red) shows a 
        distinct weakening of the broad H$\alpha$ component in the AGN (see text) with respect to earlier spectra. The earlier LAMOST spectrum (blue) taken on Apr 27, 2014, seems to correspond to the earlier SDSS spectrum of 2005 (black), but is of  lower quality.          }
   \label{fig:GNTJ143445}
   \end{figure*}

 A couple of cases deserve a short comment (see Figure \ref{fig:GNTJ143445}).  
 %Gaia15afx is a little noise, is the blue size reliable ?
 In Gaia15afx, the 3 available spectra all show clear emission lines, mainly (from red to blue) [SII], [NII], H$\alpha$, [OIII] 5007 
 and H$\beta$ (emission over an underlying absorption). The LAMOST spectrum taken on Jan 12th, 2014, although of somewhat lower S/N, 
 shows the same line intensities but a much bluer continuum compared  to  the two SDSS spectra taken in March 2002.   
 It is difficult to say whether this change is related to the Alert, which occurred much later (May 23, 2015) and no later spectrum is available either. 
GNTJ143445+3328, initially classified as a galaxy, is in fact a 2MASX AGN with a redshift of 0.197, and has associated spectra at four different epochs. But one of the SDSS 
spectra (the one of March 2010) corresponds in fact to a different object at a separation of 2.98$\arcsec$ and with a redshift of 0.246, 
and belongs therefore to a background galaxy SDSS J143445.33+332823.5. The spectra of the AGN taken by SDSS on May 1st, 2005 and by 
LAMOST on April 27th, 2014 show significant emission lines, in particular a broad H$\alpha$ component, compared to the later 
LAMOST spectrum of May 24th, 2017 (red tracing in Fig. \ref{fig:GNTJ143445}), which was taken about half a year later than the Gaia alert, 
and where the broad H$\alpha$ component and [OIII] have significantly weakened, therefore suggesting a change in the AGN properties (a Changing Look Quasar, see next section). 
 
\subsection{Active Galactic Nuclei (AGN) and Changing Look Quasars (CLQ)}

 In addition to the AGNs discussed above, there are 30 Gaia alerts and 68 GNT targets with spectra classified as ``QSO" 
in LAMOST or SDSS\footnote{The SDSS class QSO comprises two subclasses, either ``Broadline", or ``Starburst/Broadline", the latter corresponding more to lower luminosity objects where the broadline component is not dominant} , with single epoch spectra only. None of them is coincident with, or close to,  the date of the Alert, so that no clue can be given on the cause of the Alert (either a SN, or intrinsic variability of the AGN, or a Tidal Disruption Event).  
But  twelve Gaia alerts and twenty-three GNT targets have two or more spectra classified as ``QSO" in these surveys, and are therefore more interesting to consider in details.   
 For a given object, we have normalized the spectra in flux to each other at the red end (where atmospheric dispersion is minimal), 
  to reveal possible spectral changes. In many cases, no significant changes have been noticed.    
  Taking into account a probable $10\%$ (or sometimes even larger) uncertainty in the blue fluxes in the LAMOST spectra, we consider, in this work, principally   
 changes in the broad emission lines to qualify for a ``changing look" quasar, as described in \citet{mcleod16}, 
 while  changes in the continuum  are considered as an additional, less reliable argument.
  
\begin{figure*}
  % \centering
   \includegraphics[width=3.4in]{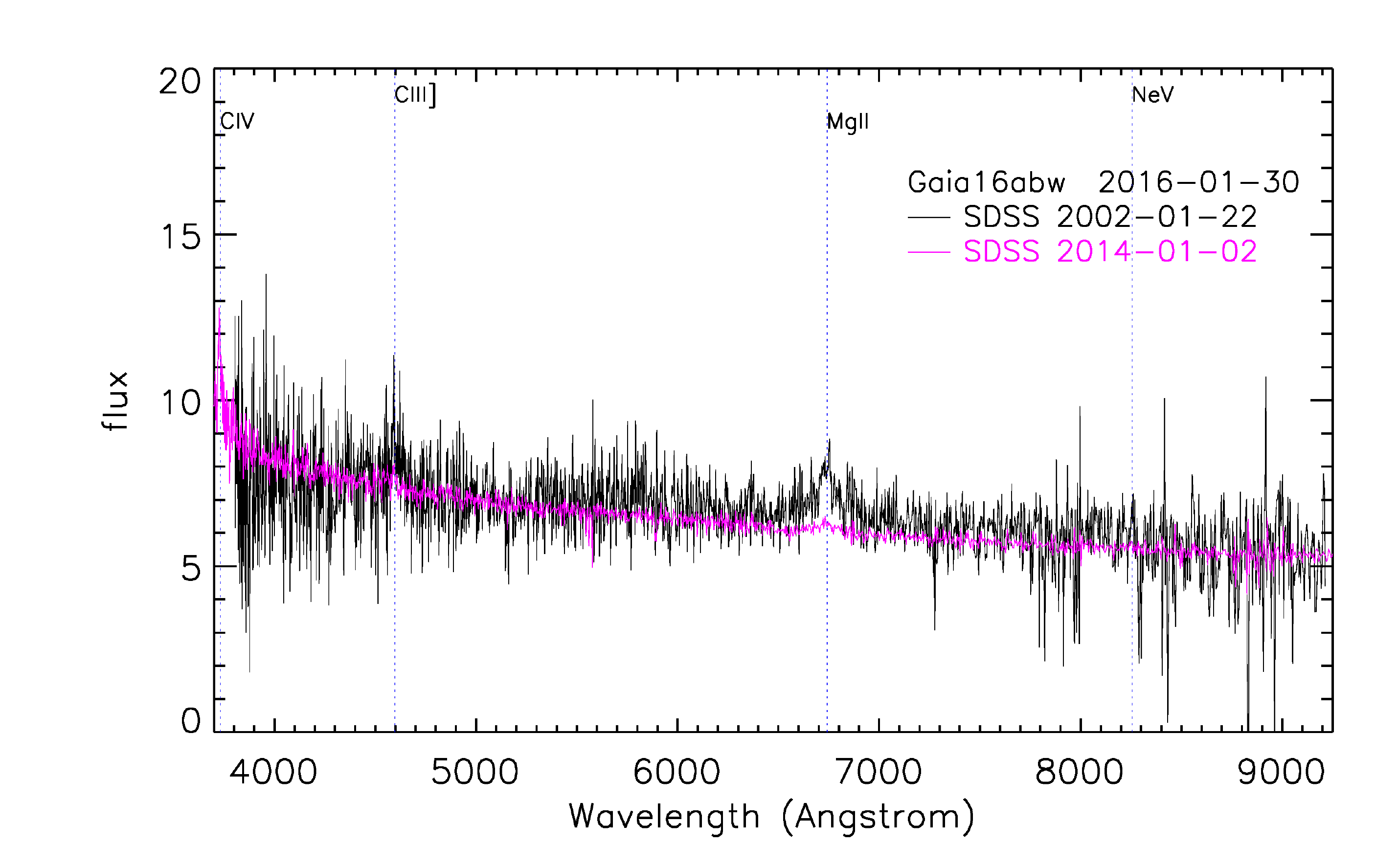}
   \includegraphics[width=3.4in]{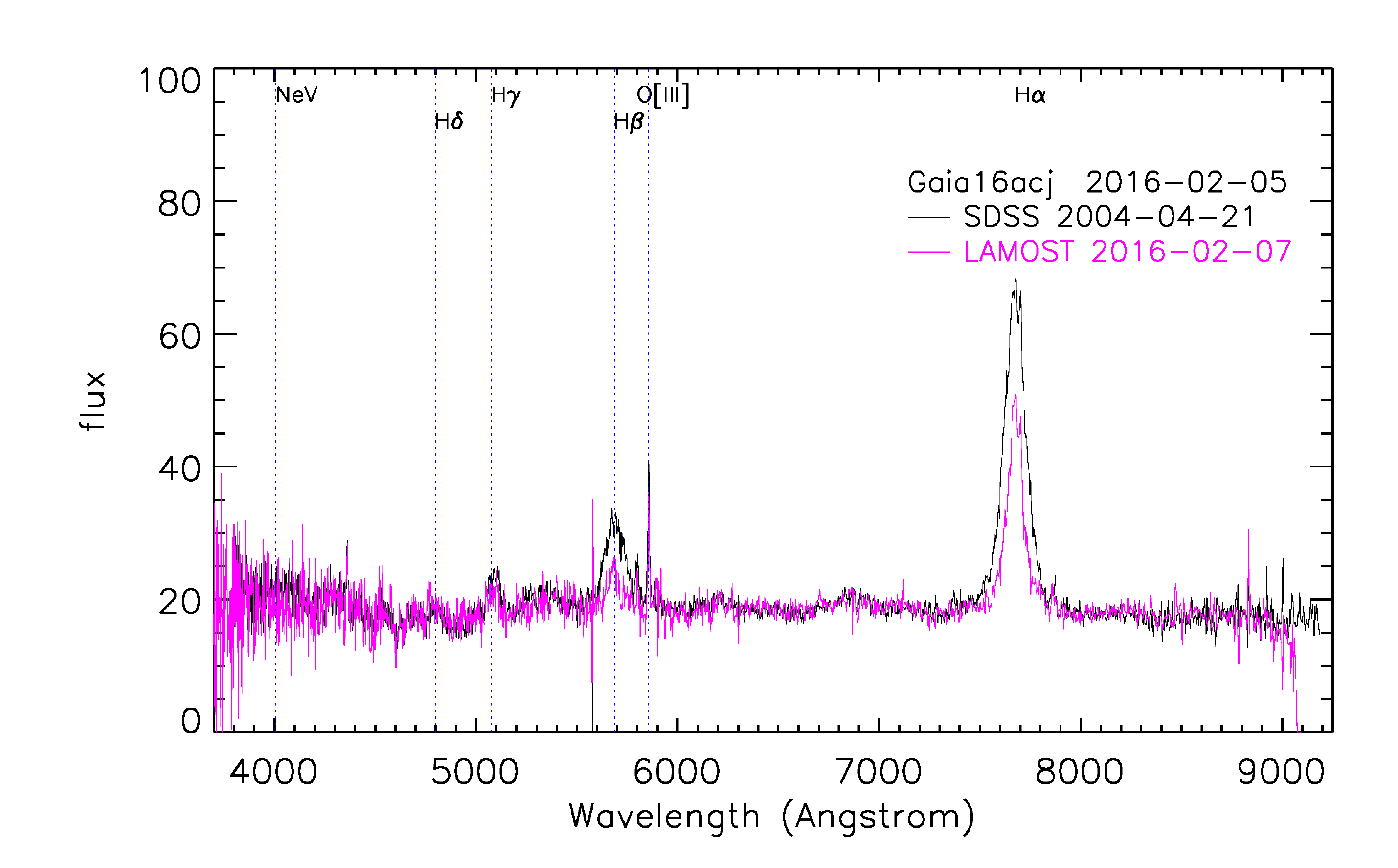}
   \includegraphics[width=3.4in]{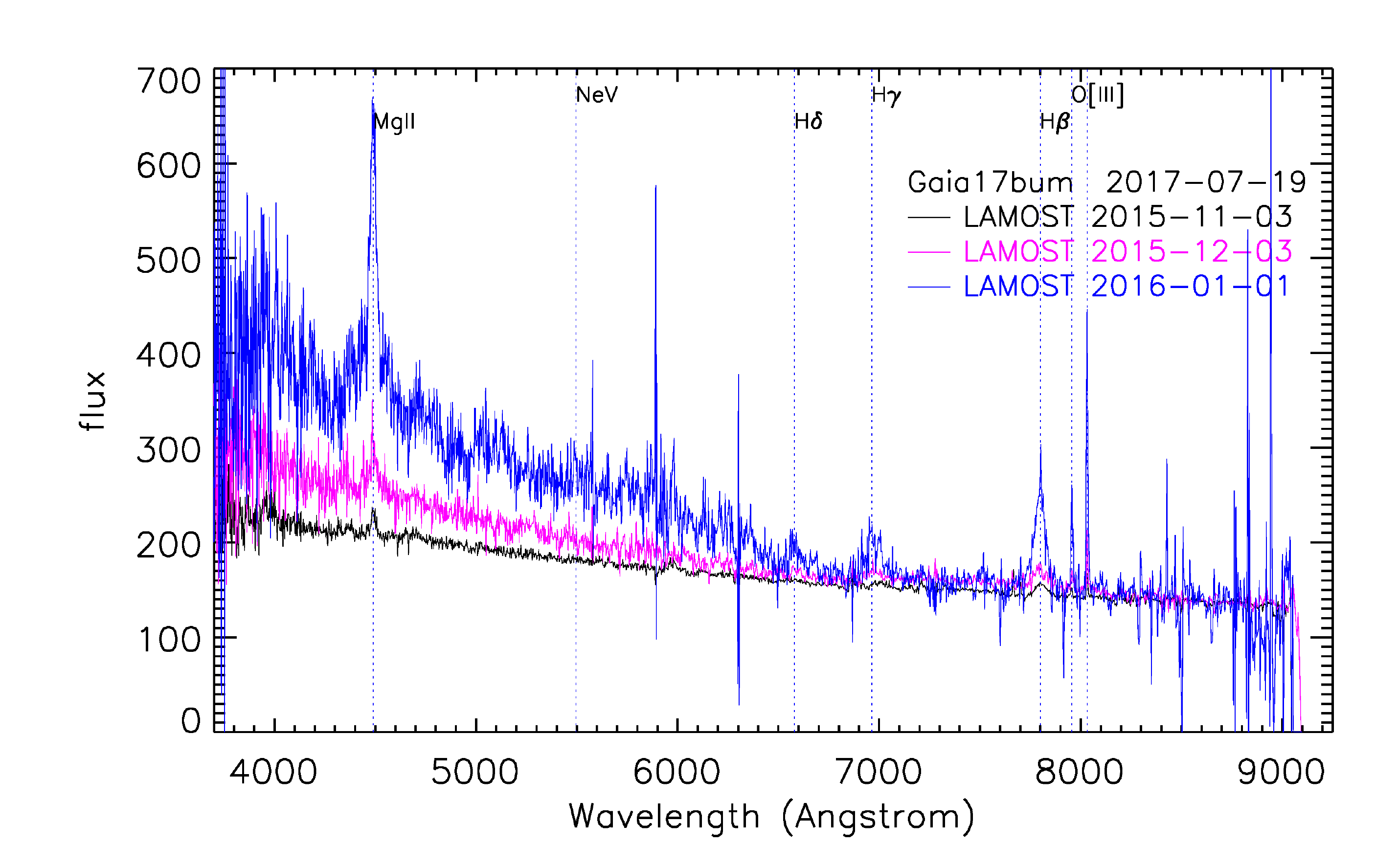}
   \includegraphics[width=3.4in]{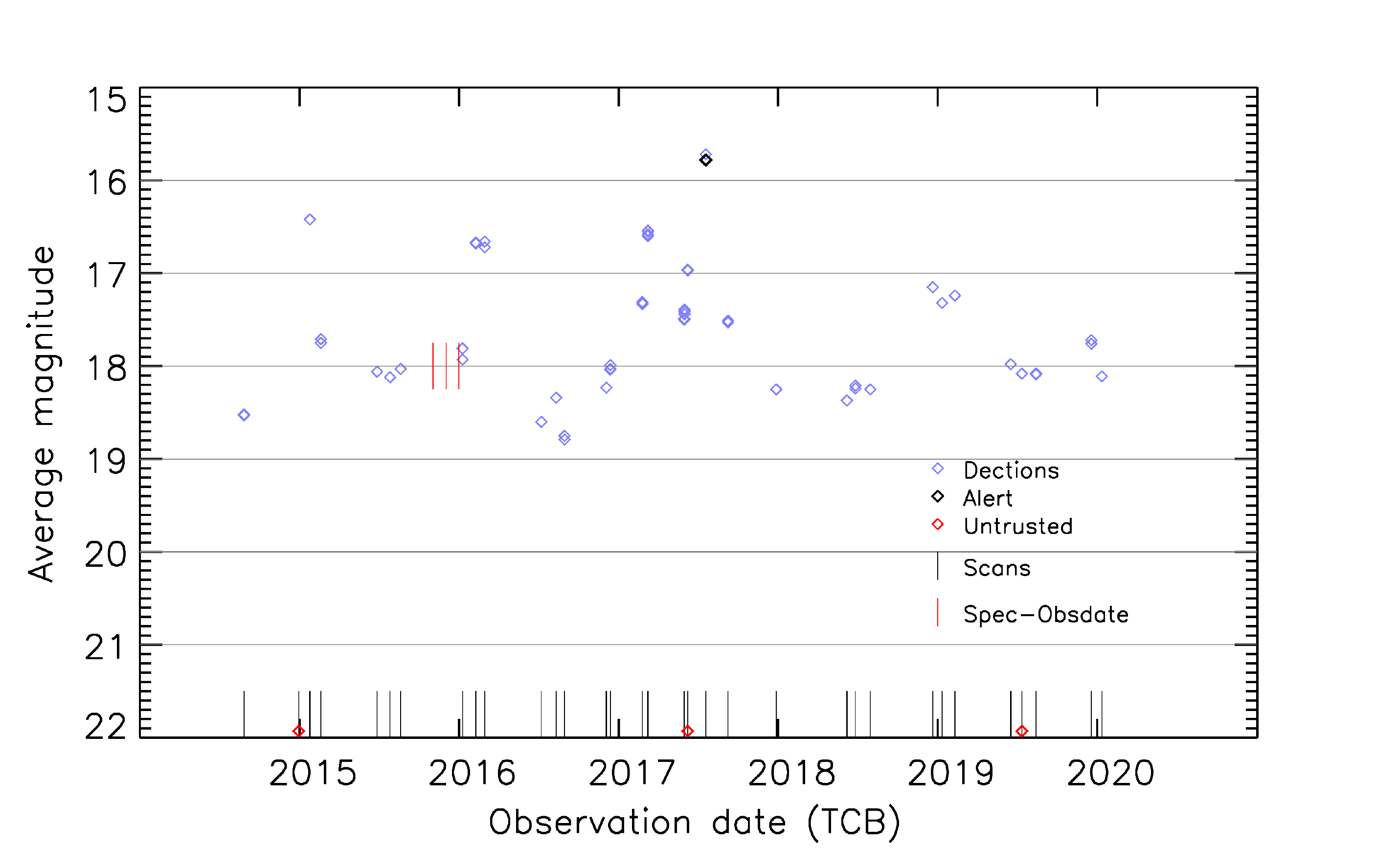}
       \caption{Three Gaia alerts of QSOs showing some changes in the broad emission lines and/or in the continuum, 
        with multi-epoch spectra normalized at the red end. The light-curve of Gaia17bum is from Gaia Alerts, with dates of 3 spectra marked by red vertical lines.               }
         \label{fig:Gaia_QSOs}
   \end{figure*}

\begin{figure*}
   \includegraphics[width=3.4in]{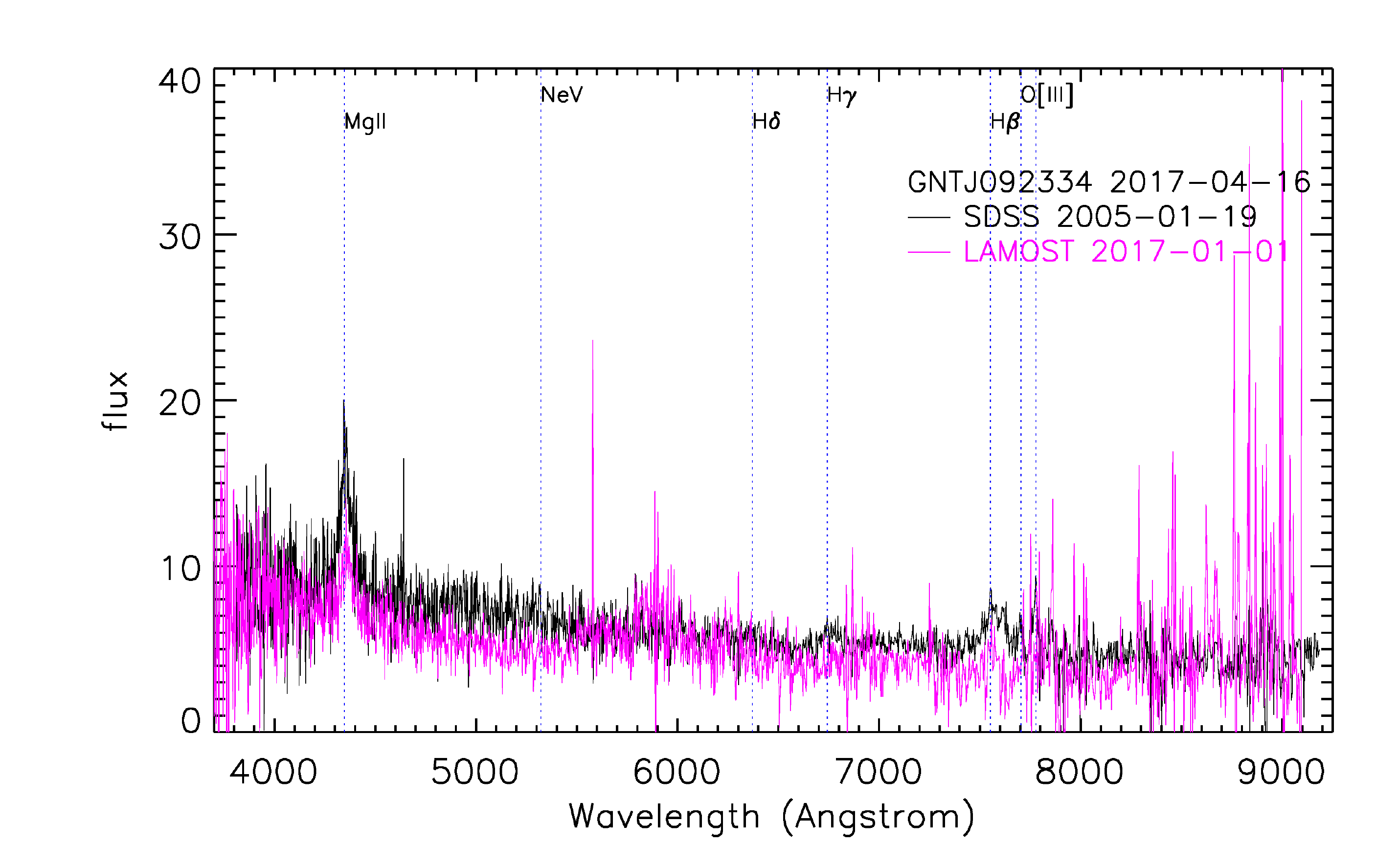}
   \includegraphics[width=3.4in]{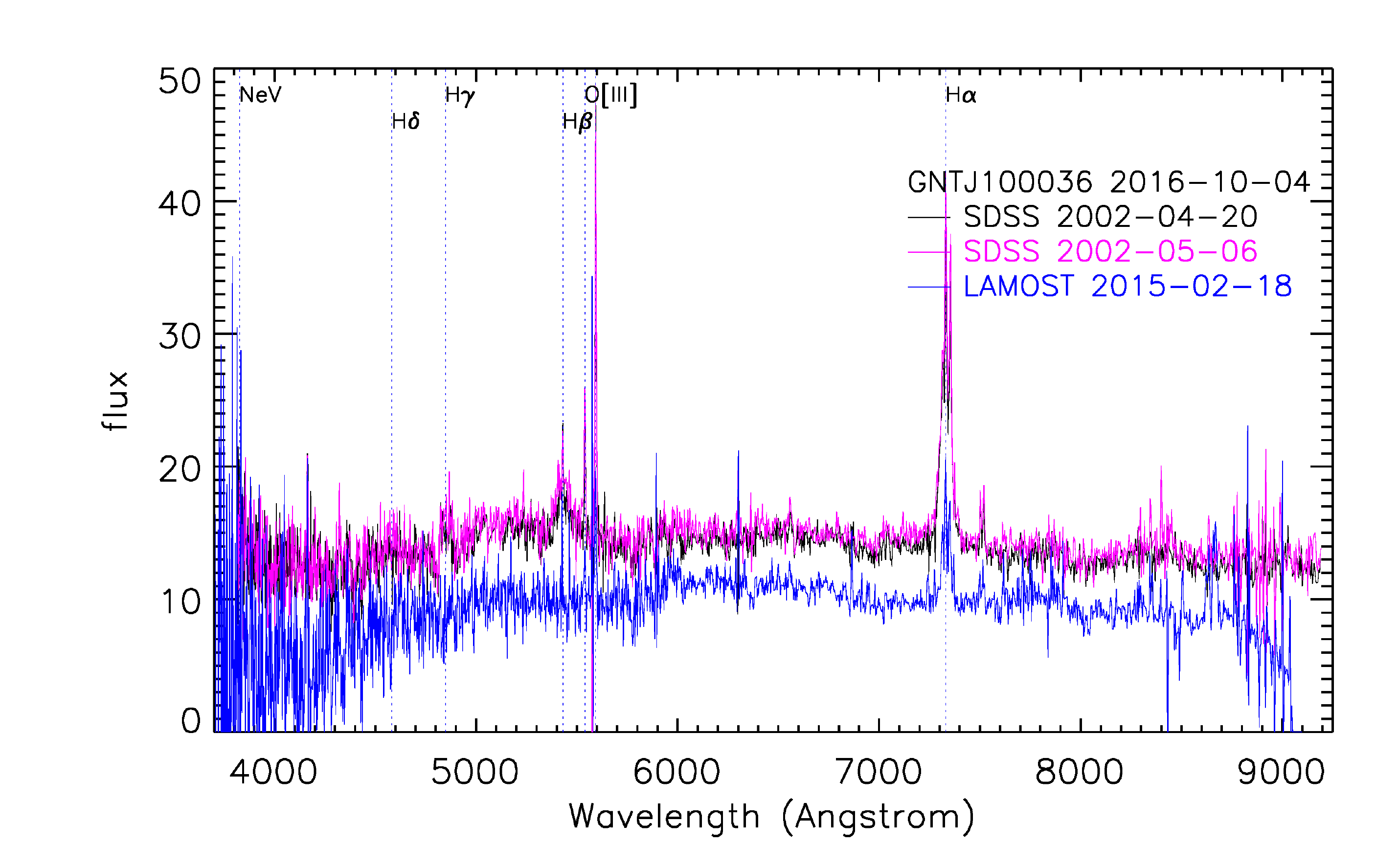}
   \includegraphics[width=3.4in]{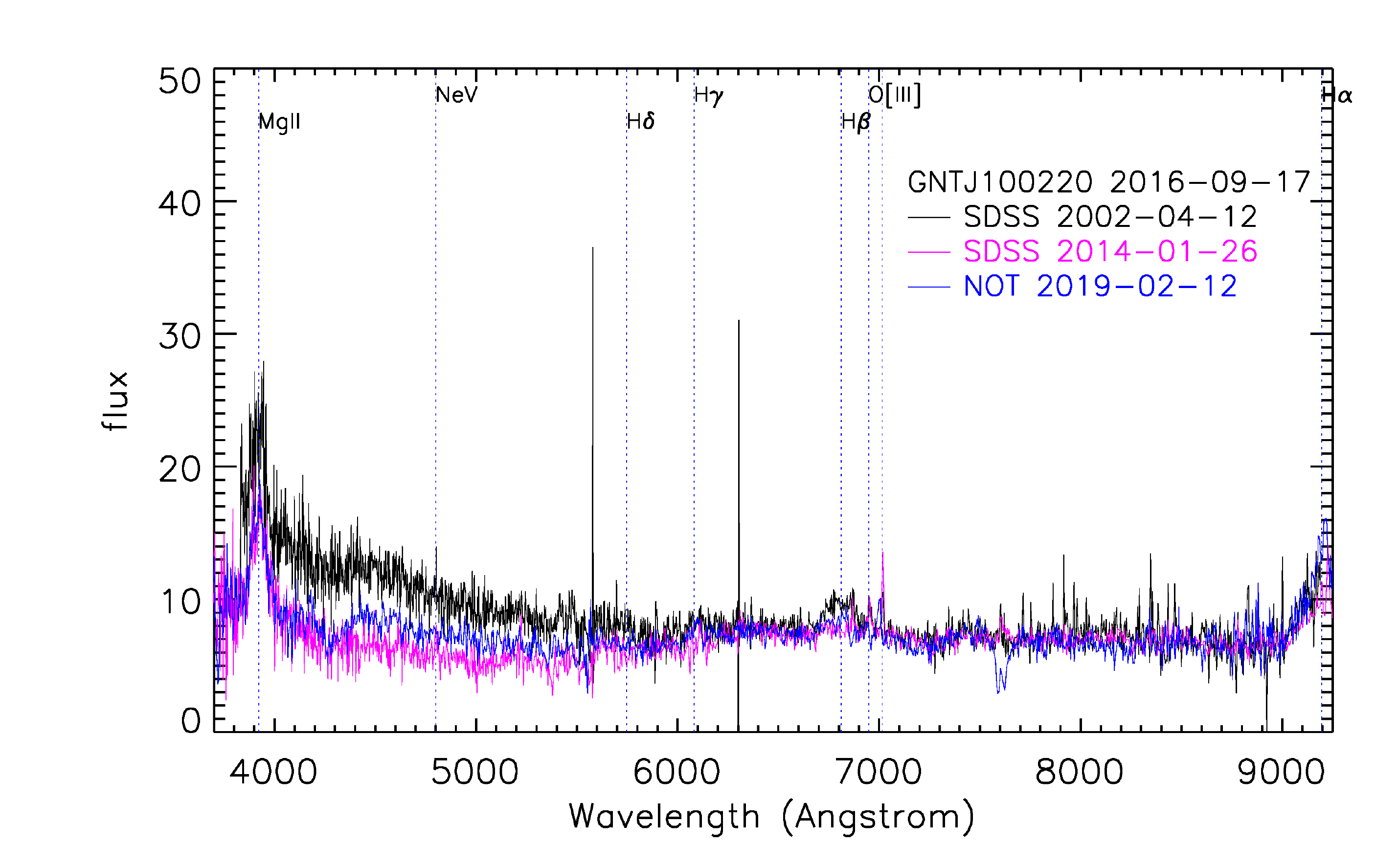}
   \includegraphics[width=3.4in]{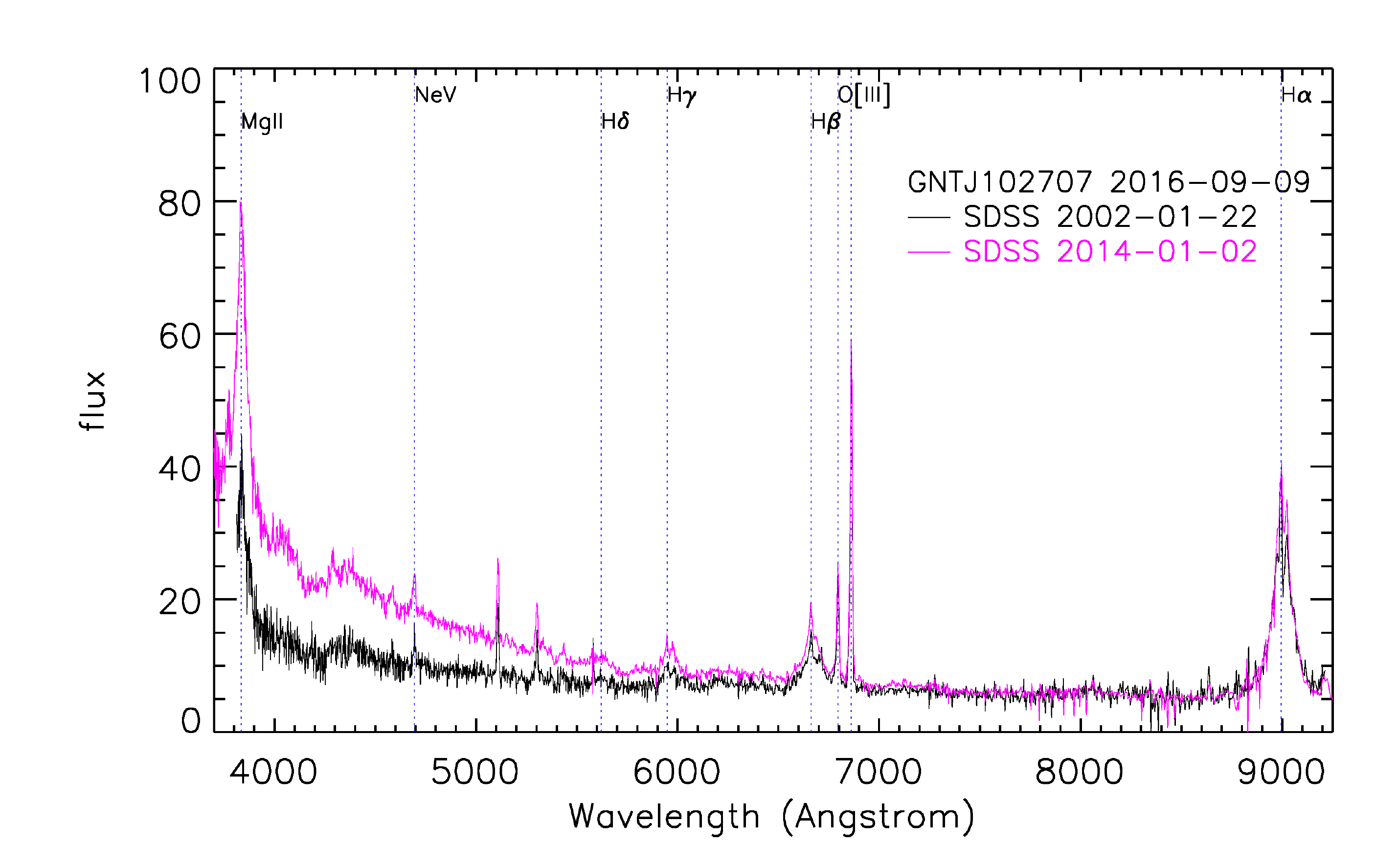}
   \includegraphics[width=3.4in]{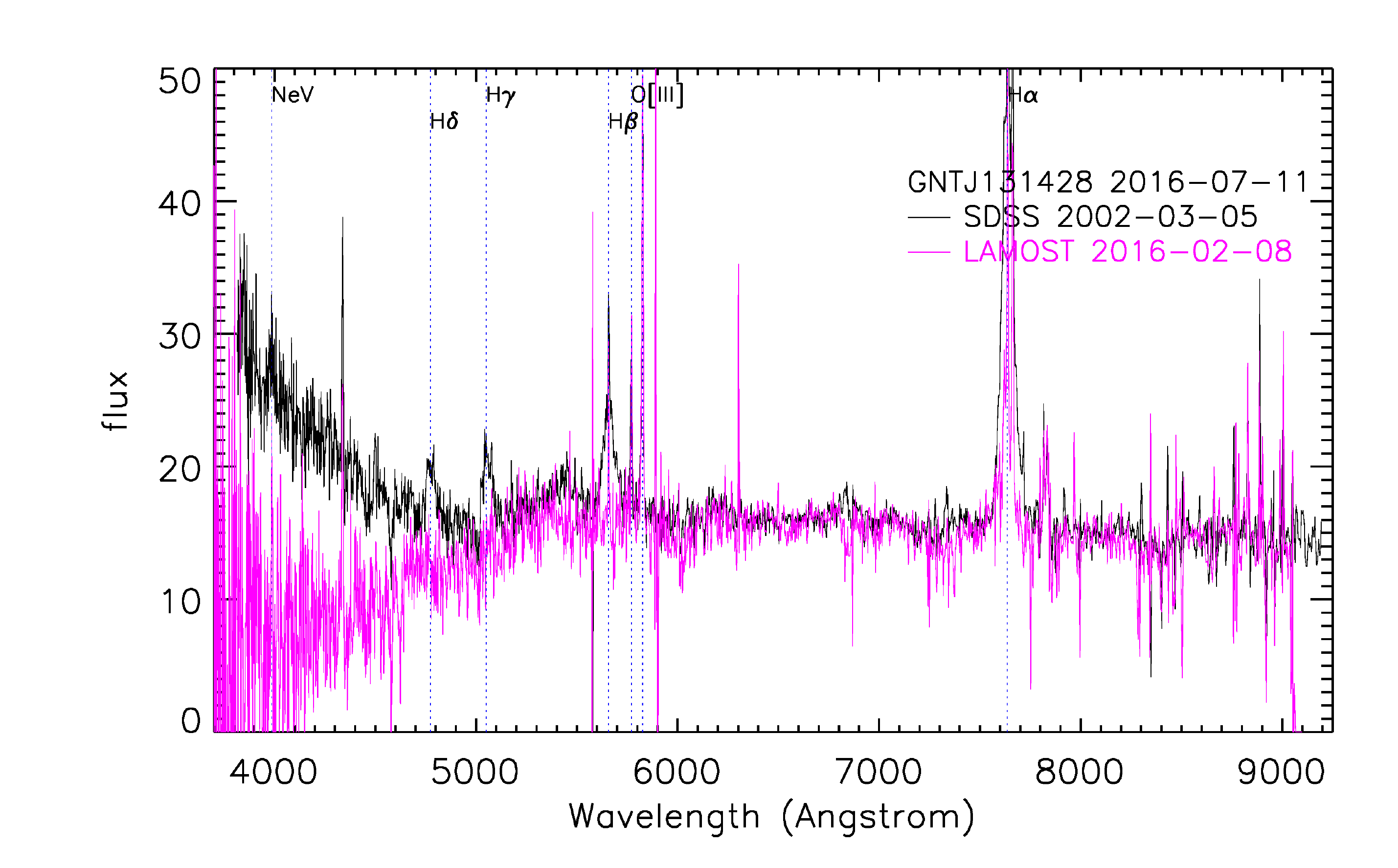}
   \includegraphics[width=3.4in]{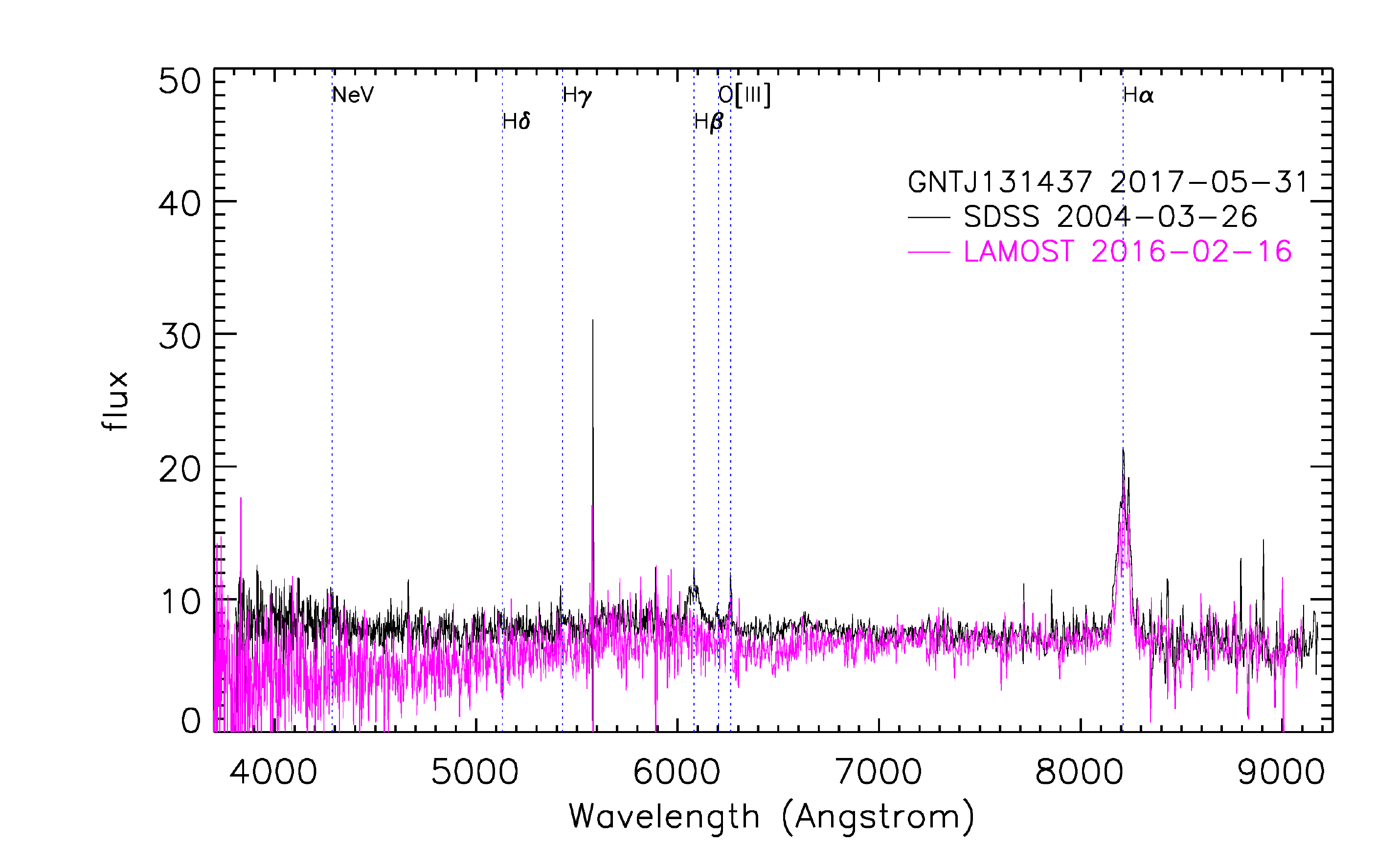}
   \includegraphics[width=3.4in]{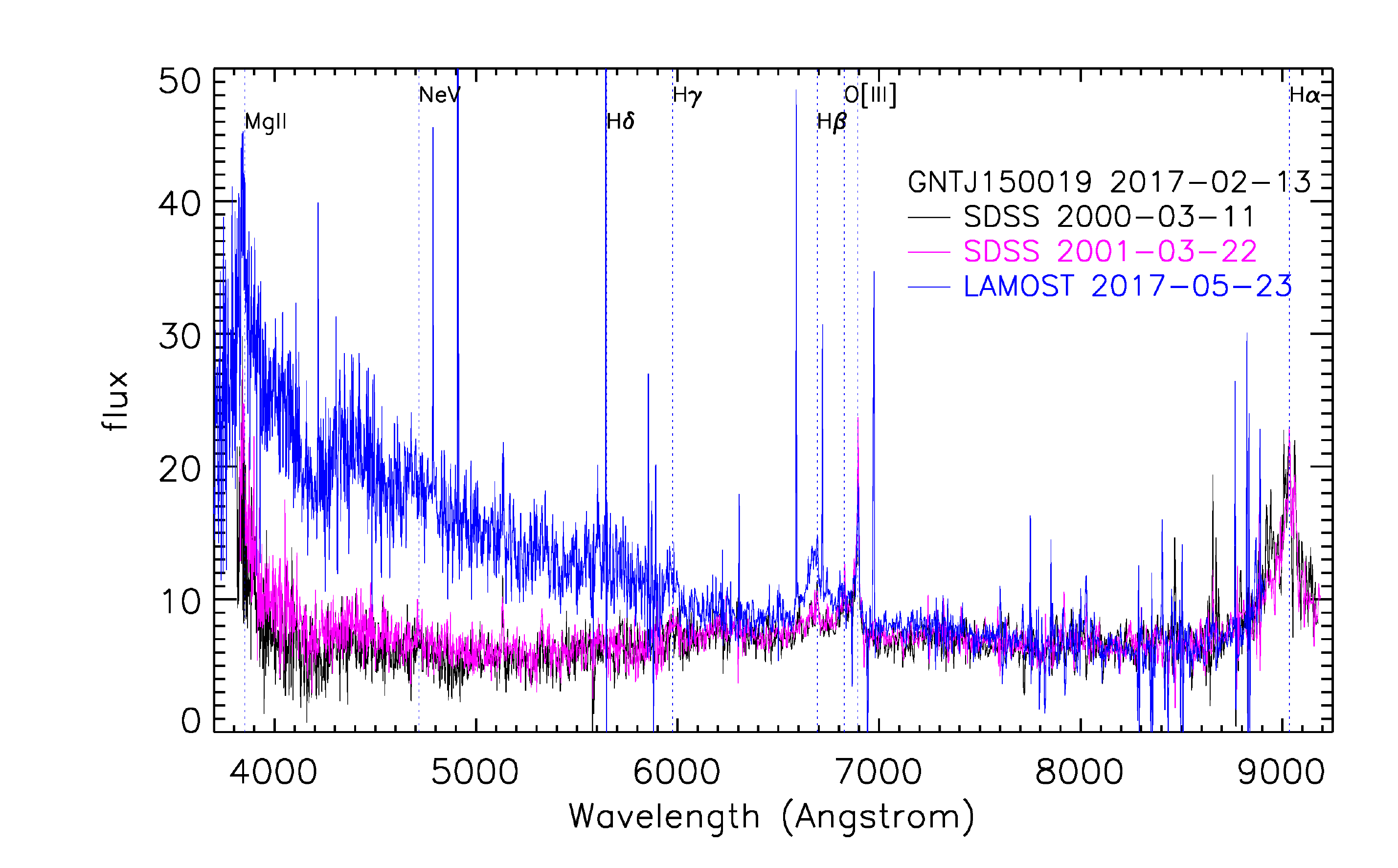}
       \caption{Seven QSOs from the GNT list with multi-epoch spectra (normalized at the red end) and showing some changes in 
       the broad emission lines and/or in the blue continuum. For GNTJ100036+5116, the spectra are shown not normalized to each other,  for clarity in the figure.
              }
         \label{fig:GNTlist_QSOs}
   \end{figure*}

A dozen of targets (out of 35) show some changes in either the emission lines or in the intensity of the continuum in the blue,  or both, as illustrated  in Figure \ref{fig:Gaia_QSOs} and \ref{fig:GNTlist_QSOs}. All of them are  classified as QSOs (except Gaia16abw which is in fact a blazar). 
 For instance, the 2 SDSS spectra of Gaia16abw taken at 12 years interval show a difference of more than a factor of 2 in the continuum, with corresponding changes in the intensity of the emission lines, but a general shape otherwise totally similar: 
  the normalization then reveals a significant decrease of the intensity of the MgII 2800$\AA$ emission line between  
  2002 and 2014, its EW changed  by more than a factor of two, see Figure \ref{fig:Gaia_QSOs}, and a similar decrease of the CIII] 1909 line. With the high redshift of this blazar (z = 1.41), 
 it is unlikely that this change is due to the change of entrance fiber, from 3" to 2", from DR9 onwards, so it should be considered as real. The Gaia data, between early 2015 and end of 2019,  show indeed its G magnitude oscillating between 19.5 and 15 (the Alert being issued on a high point in Jan. 2016) and the spectra are consistent with the lines responding to a change in the continuum. 

Gaia16acj, a QSO at redshift 0.17, shows significant changes in the broad line H$\beta$ and H$\alpha$ intensity, and the LAMOST spectrum was 
taken only two days after the Gaia alert (Feb. 5th, 2016, pointing to a brightening of the QSO): the smaller equivalent width of the emission
 lines seen in the LAMOST spectrum may thus indeed be due to the brightening of the continuum, before the lines could respond to this change.  

Gaia17bum, observed at three different epochs, shows a significant increase in the MgII,  H$\delta$, H$\gamma$ and H$\beta$ emission 
intensities between the first spectrum of Nov. 3rd, 2015 and 
 the last one (Jan. 1st, 2016, blue line in Fig. \ref{fig:Gaia_QSOs}) as well as a continuum getting bluer. 
Although the Gaia Alert was emitted  about 1.5 years later than this last spectrum, the Gaia light curve showed some previous peaks (not alerted for) closer to the spectrum, 
for instance on Feb. 8th, 2016.  But it apparently remained  in the low state between June 2015 and the last low point on Jan. 9th, 2016 (that is,  after the last LAMOST spectrum of Jan. 1st, showing significant increases), before getting high on Feb. 8th. There seems thus to be a lag between 
photometric and spectroscopic changes, the value of which is however impossible to assess here with so irregular data points  
(no Gaia points are available at dates close to the other two, earlier  LAMOST spectra). The interval between significative changes, as seen from its Gaia light curve, seems to be some weeks or months, but not years. \\
Two other Gaia Alerts show some weaker changes:  Gaia16ack  in the H$\beta$ line, and Gaia17czm in the CIV line, but the odd shape 
around 5900$\AA$ (due to the  LAMOST red/blue band splitting) makes the estimate of the continuum rather uncertain there.   
 
 \begin{table*}
\footnotesize % \small
\caption{Changing look AGN and QSOs.}             % title of Table
\label{table_cl}
\centering
\begin{tabular}{l c c c c c c c c c c c}
\hline\hline    
Name          &    Ra          &        Dec  &  Pre Gmag  & Alert Gmag  & Epoch  & Redshift & Emission Line & Continuum Shape \\
\hline
Gaia16abw & 158.46428 & 60.85202  &          &  15.66   & 2    &   1.40871 & decrease & \\
Gaia16acj   & 196.21654  &11.13351  &          &  18.01   & 2    &    0.16914 & decrease & \\
Gaia17bum & 18.209680  &32.13809 & 17.31&  15.75   & 3   &    0.60384 & increase & bluer \\
\hline
\hline
 Name         &             Ra   &      Dec  &  Pre Gmag & Gmag Peak   & Epoch  & Redshift & Emission Line & Continuum Shape \\
 \hline
GNTJ143445.35+332820.56 & 218.68896&  33.47238 & 19.21 & 18.85  &    3  &   0.197561 & decrease & \\
 \hline
GNTJ092334.70+281526.86  &140.89458  &28.25746 &  19.61 & 19.17    & 2  &   0.55305 & decrease & \\
GNTJ100036.45+511652.91  &150.15186  &51.28136  & 19.67  &18.69    & 3   &  0.11668 & decrease & \\
GNTJ100220.18+450927.30  &150.58410  &45.15758  & 19.32  &18.60    & 3   &  0.40078 & decr/incr & redder/bluer\\
GNTJ102707.45+602633.62  &156.78105  &60.44267  & 18.92  &18.42    & 2   &  0.37012 & increase & bluer \\
GNTJ131428.09+054307.34  &198.61705  &5.71871   & 19.63  &18.87     & 2   &  0.16331 & decrease & redder \\
GNTJ131437.60+142503.90  &198.65667  &14.41775  & 19.91  &19.32    & 2  &   0.2506 & decrease & redder\\
GNTJ150019.08+000249.02  &225.07951  &0.04695   & 19.08  &18.39     & 3   &   0.37637 & increase & bluer \\
\hline
\end{tabular}
\end{table*}
 
  In the QSOs from the GNT list with multi-epoch spectra, about ten of them display changes, seven of which are found 
  to be strong, see Figure \ref{fig:GNTlist_QSOs}.   This larger fraction is obviously due to the target selection in this list (``Nuclear Transients"). 
  In GNTJ092334+2815 and GNTJ100036+5116, the broad H$\beta$ line disappeared in the LAMOST spectra compared to 
  the SDSS ones taken about 10 years earlier, while the H$\alpha$ line also looks weaker in GNTJ100036+5116 (but it falls outside the red   end of the spectrum of GNTJ092334+2815 and can thus not be checked for that one). 
    
The quasar GNTJ100220+4509 was found by Gaia to have increased its G magnitude from 19.32 to 18.64 \citep{kostrz18} on Sept. 17th, 2016,
thus after the last SDSS spectrum was taken.
 The first SDSS spectrum, from Apr 12th, 2002, shows a classical broad H$\beta$ component with weak [OIII] lines. 
  Only the blue side of the broad H$\alpha$ component is visible at the edge of the spectrum because of its redshift z= 0.401. 
  The second SDSS spectrum, obtained on 2014 Jan 26th, thus still before the Gaia Alert, shows that the broad H$\beta$ 
  component has disappeared. 
%, and the concomitant broad H$\alpha$ component has weakened, a narrow component being present above it. 
  The blue continuum has slightly weakened also. 
  We have obtained a new, ground-based spectrum in Feb. 2019 with the Nordic Optical Telescope (NOT), which shows that the H$\beta$ broad component has 
  reappeared, in an interval of therefore 5 years, see Fig. \ref{fig:GNTlist_QSOs}. 
We are therefore here in the presence of a ``changing look quasar", as described in McLeod et al. 
(\citeyear{mcleod16})\footnote{Note that this object is the only one in common with their sample}, showing both disappearance and reappearance of the broad Balme component. 
The fact that the observing dates are not coincident 
  between Gaia and ground facilities does not allow to decide whether the reappearance of the broad Balmer component  
  is directly linked to the brightness increase detected by Gaia in 2016 and what would be the time delay between 
  photometric and spectroscopic changes.  
  %\textcolor{blue}{We are therefore here in the presence of a "changing look quasar" as described in McLeod et al.  (\citeyear{mcleod16}).
       
  GNTJ102707+6026 shows a clear increase in the blue continuum from 2002 to 2014, with no changes in the 
  Balmer lines. This continuum increase also better reveals the strength of the FeII multiplets redwards of 
  the MgII line (e.g. \citealt{grandi81} for reference), but the alerting date (Sep. 9th, 2016) is not covered by the spectra.
  
  For GNTJ131428+0543,  the Balmer emission lines (except H$\alpha$) have disappeared, 
  and the blue continuum dropped, between the SDSS spectrum of March 2002, and the LAMOST spectrum of 
 Feb. 8th, 2016, a few months before Gaia alerted on a change on July 11th the same year. 
 For GNTJ131437+1425, the situation is almost the same, but the drop in blue continuum  is only marginal. 
  
  Another case, with opposite changes, is detected in GTNJ150019+0002:  a broad H$\beta$ component, weak or absent in the  two SDSS 
  spectra of 2000 and 2001, has appeared on the LAMOST spectrum of May 23rd, 2017, that is  a few months after the Gaia detection of 
  Feb. 13th the same year, while no change is noticed in the H$\alpha$ line (which sits however at the red edge of the spectrum). 
  The blue continuum has also increased, and there is evidence that the FeII multiplets redwards of the MgII line have increased also. 
  We have thus here another clear case of ``Changing Look Quasar", possibly associated with the magnitude change noticed by Gaia.
We list the information for these ten CLQs as well as the one CLAGN in Section 3.2 in Table \ref{table_cl}.

We also note here three cases with relatively weaker changes:  GNTJ121613+5242 shows a weak change in the shape of the 
H$\delta$ and H$\beta$ lines, GNTJ150149+2830 in H$\beta$ only, while GNTJ232841+2248 displays changes in the shapes of 
both H$\beta$ and H$\alpha$. 
 
 From the few cases seen here, it seems thus that an increase in H$\beta$ line intensity is coming together with the 
 increase of the blue continuum, and vice-versa. While this small sample is not statistically sufficient, it goes along the findings of 
\cite{mcleod16} and almost doubles their sample.  All the objects investigated here, like theirs, have first been selected because of 
photometric variability. On the contrary, \cite{Yang18} have recently looked for CLQ's by searching directly for repeat spectra in the 
LAMOST or SDSS databases and found another 21 cases, confirming the trend of bluer when brighter, but surprisingly, there is no object 
in common with our sample! The reason for this (selection effects ?) needs to be investigated further. Similarly, while this paper was with 
the referee, \cite{Graham2020} published a list 73 CLQ's selected by photometric variability in the  CRTS survey of known quasars, and having 
confirmed spectroscopic changes: none of them is in common with our list, but a full cross-check of the Gaia Alerts with CRTS remains to be done. 
 What is really needed in the future, is a closer monitoring in time, to establish the possible link, and time delay between the changes 
in  the magnitude, and  the observed spectral changes.  

% \begin{landscape}
%\end{landscape}

\subsection{Various}

\begin{figure}
  \centering
   \includegraphics[width=3.3in]{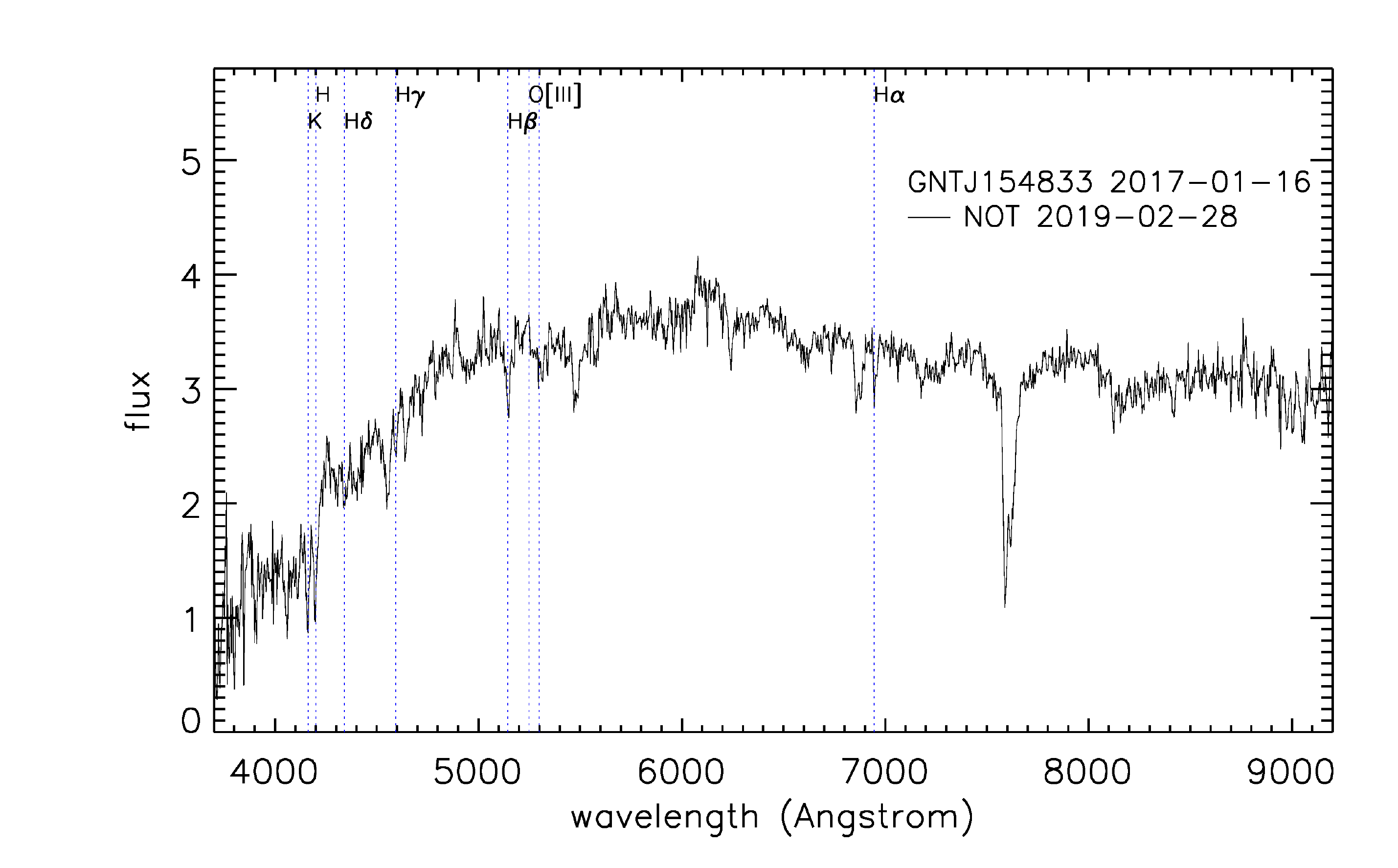} 
       \caption{Spectrum of the galaxy GNT J154833-0017 obtained at the NOT in Feb. 2019.}
         \label{fig:GNT1548}
   \end{figure}

Eight objects (7 Gaia alerts and 1 GNT)  have LAMOST spectra whose (automatic) classification was marked as ``Unknown", 
essentially because of too low a S/N ratio, but none of the spectra was coincident in time with the Alert. A closer examination of them allows to extract further information. 
For GNT154833-0017, alerted on Jan.16, 2017, a much earlier (2012) LAMOST spectrum exists, but its low S/N did not allow to 
conclude about the nature of the galaxy. A new, NOT spectrum was obtained by us on Feb. 28, 2019, showing a galaxy at redshift 0.062 with no emission lines (see Fig. \ref{fig:GNT1548}). 
This WISE galaxy is catalogued at mag g=17.6, without redshift,  and is therefore the host of the event, but no clue can be given on the nature  of that event.  
 Note that another WISE galaxy is found about 1' NE with z = 0.061 and thus belongs to the same group. 

Gaia17cdi was alerted as a candidate Ia SN on August 22, 2017, but the LAMOST spectrum dates back to Jan. 29, 2014, 
and shows possibly H$\alpha$, [NII] and [SII] in emission, indicating in fact a galaxy with a redshift  of $\sim 0.073$,   which needs however confirmation. 
Nothing can be concluded from the LAMOST spectra nor from the extra-galactic databases for Gaia17cxa (no catalogued host),  
or Gaia18bxf (coincident with a 2MASX host, but no redshift available). 
For Gaia17djf, the much earlier (Oct. 2013) LAMOST spectrum shows in fact clearly H$\alpha$ and [NII] in emission, or Na in absorption, 
thus giving a redshift of 0.048, consistent with the one of 0.049 found in the databases for a 2MASX  galaxy (some other, 
broader features appear further in the red, possibly related to another, background object, which needs confirmation). 
For Gaia18adz, the 2MASX host has no measured redshift in the databases, but the LAMOST spectrum displays a possible H$\alpha$ in emission, suggesting a redshift 
of $\sim 0.19$, but this needs confirmation (the corresponding absolute magnitude for the SN candidate would be too high). 
For Gaia18aov, the LAMOST spectrum taken several years earlier (June 2012), although classified as ``Unknown",  
shows clearly H$\alpha$, [NII] and [SII] in emission, indicating a redshift of 0.0118, giving a plausible,  absolute magnitude of -15.3  for the SN candidate. 
This is in line with the redshift of 0.011826 found in the NED database for a very closeby galaxy, CGCG 275-003, thus probably the host of the SN.  
Finally, Gaia18bzv, a possible SN candidate in the galaxy UGC2536, has been identified as a type Ia SN (SN2018eqq) at z= 0.016 
(corresponding to the redshift of UGC2536) and the LAMOST spectrum of Nov. 2012, although of low S/N and classified as ``Unknown", 
shows a strong Na, and weaker MgI or H$\alpha$ absorptions at this redshift, and has therefore recorded the spectrum of the host galaxy.  

\section{Conclusions}

   \begin{enumerate}
      \item We have compared over 6300 Gaia Alerts, complemented with 482 Gaia Nuclear Transients (GNT), with 
 objects whose spectra had been recorded in the LAMOST  spectroscopic survey and/or  in  
  the SDSS DR15 survey.

      \item We have identified, among the Gaia Alerts, 14 stars, 157 galaxies and 30 QSOs with single epoch spectra, 
          plus 42 galaxies and 68 QSOs among the GNT candidates. \\
          In addition to those, some  objects have multi-epoch spectra (without distinction between LAMOST or SDSS): we identify  12 stars  (plus a mis-identified one which is in fact a BL-Lac object), 12 QSOs and 49 galaxies in the Gaia alert sample; 
   and 13 galaxies and 23 QSOs in the GNT sample. \\
   The relatively small number of identifications (compared to the total number of input objects) is partly due to the 
   limiting magnitude and sampling of spectral observations, and to the fact that both LAMOST and SDSS 
   cover only the northern part of the sky while Gaia is an all-sky survey.
   \item Among the Gaia Alerts having LAMOST or SDSS spectra labelled as ``STARS", most  of them (13) are cataclysmic variables, some of which display double-peaked emission lines.   
    \item For most galaxies, the alert  was probably caused by a SN explosion, but no confirmation can be given here as the spectra do not coincide in time with the Alert.  
      In only two cases, GNT J170213+2543 and Gaia17aal, are the ground-based spectra contemporary to the 
      alerting date and their spectra reveal residuals from a SN: these two, previously unclassified, candidates are now confirmed here as type Ia SNe.  
     \item Ten quasars with multi-epoch spectra show significant changes in the broad emission lines, with sometimes 
 also changes in the slope of the blue continuum. 
A few candidates with weaker changes are also identified. 
  These changes  
qualify these objects  as  ``Changing Look Quasars",  as described in  \cite{mcleod16},  
 almost doubling their sample, and confirm the trend of the object getting bluer, with a stronger broad Balmer component, when getting brighter. 
What is needed to better understand the physics underlying those changes are spectra taken closer to the photometric Alert and at regular intervals to quantify the time scales and eventual time-delays between photometric and spectroscopic changes.   
        
   \end{enumerate}

% Acknowledgements

 \acknowledgments
This work was initiated during a visit to SWIFAR (in the frame of the Chinese-French LIA ``Origins" program), whose hospitality and 
support are gratefully acknowledged by MD. 
ZYH thanks Dr. Wei Zhang, Zhongrui Bai and Dongwei Fan for helpful discussions. ZYH and XWL have been partially supported by
 the National Key Basic Research Program of China 2014CB845700. 
 TMZ is supported by the NSFC 11633002.
We acknowledge the use of data from the ESA Gaia Satellite, DPAC and the Photometric Science Alerts Team (http://gsaweb.ast.cam.ac.uk/alerts). \\
 The Guoshoujing Telescope (the Large Sky Area Multi-Object Fiber Spectroscopic Telescope LAMOST) is a National Major 
 Scientific Project built by the Chinese Academy of Sciences. Funding for the project has been provided by the National 
 Development and Reform Commission. LAMOST is operated and managed by the National Astronomical Observatories, 
 Chinese Academy of Sciences. \\
We used data from  the Sloan Digital Sky Survey (SDSS) public releases. Funding for the SDSS has been provided by the Alfred P. Sloan 
Foundation, the U.S. Department of Energy Office of Science, and the Participating Institutions (details on the SDSS web site at www.sdss.org). \\ 
This research has made use of the SIMBAD database, operated at CDS, Strasbourg, France, and of the NASA/IPAC Extragalactic Database (NED), which is funded by the National Aeronautics and Space Administration and operated by the California Institute of Technology. \\ 
We thank an anonymous referee for valuable comments which helped to improve the manuscript. \\

%  Using BibTeX  (Name-Year style)
\nocite{*}
\bibliographystyle{spr-mp-nameyear-cnd}  %% BibTeX style
 \bibliography{zyhuo_ms}                %% BibTeX data

\begin{appendix}

\section{Tables}
Spectral information for the Gaia Alerts and the GNT sample are given in the Tables below. Here we only display a few lines as examples,  while  the full Tables are available on line only.
Table \ref{table:table1b} lists all the 387 spectral information available for the 285 targets in the Gaia Alerts.  
This includes the name of the object, Ra and Dec in decimal degrees, the
mean historic magnitude in the Gaia G band, the G band magnitude at the time of the alert, the date of the alert,  the date of the spectral observation, the origin of the
spectra (LAMOST or SDSS), Ra and Dec of the spectral data in decimal degrees, redshift, class and subclass of the object,  the spectral type of the object: 
emission or absorption line star/galaxy (star: ELS/ALS, galaxy: ELG/ALG; if both emission and absorption lines are present, we note E+A or A+E), and comments about the origin  of the photometric variability, if available. \\
Table \ref{table:table2b} lists all the 192 spectral information for the 146 targets in the GNT list, the columns are the same as in 
Table \ref{table:table1b}, except that the name,  
the G band medium magnitude, the G band peak magnitude,  and the date of the G band peak are taken from \cite{kostrz18}. 
Some additions/corrections  are introduced in this table since  their work, also a few more identifications and redshifts (from LAMOST, SDSS or NOT) 
as indicated in the comments, and the multi-spectra  description. \\
Full tables are  available at CDS via anonymous ftp to cdsarc.u-strasbg.fr (130.79.128.5) 
or via http://cdsarc.u-strasbg.fr/viz-bin/qcat?J/other/A+SP,

\setcounter{table}{0}
\begin{landscape}
\begin{table*}
\tiny
\caption{Spectral information for Gaia alert targets.}             % title of Table
\label{table:table1b}      
%\centering                          
\begin{tabular}{lllllllllllllll}        
\hline
\hline    
Name  &  Gaia-Ra & Gaia-Dec & Pre-Gmag &Alert Gmag     & Alert-Date & Spec-Obsdate &  Spec-Tels & Spec-Ra & Spec-Dec & Redshift &   Class & Subclass &  Type & Comments \\               
\hline    
 Gaia14aaa&      200.25961&      45.53943&   19.22&   17.32&  2014-08-30&   2004-02-20 &   SDSS&      200.25944&      45.539947&          0.035907&  GALAXY&  STARFORMING&      ELG&     SN Ia\\
 Gaia14aab&      206.91011&      55.54871&   20.00&   19.11&  2014-07-27&    2013-05-16 &  SDSS&      206.91008&      55.548800&         -0.000684&    STAR&                         sdA&      ALS&     unknown\\
 Gaia14aac&      214.66744&      56.46959&   19.55&   18.68&  2014-08-02&    2003-03-05 &  SDSS&      214.66744&      56.469549&          0.074111&  GALAXY&                             &       ALG&     unknown\\
 Gaia14aak&      213.74354&      38.19487&   19.72&   19.06&  2014-09-13&    2004-03-20 &  SDSS&      213.74355&      38.194874&          0.064287&  GALAXY&                     AGN&      ELG&     unknown\\
 Gaia14aak&      213.74354&      38.19487&   19.72&   19.06&  2014-09-13&    2016-02-09 & LAMOST&     213.743537&      38.194872&       0.063833401&    GALAXY&             Non&       ELG&     unknown\\
 Gaia14aam&      181.39832&      49.47269&   19.97&   19.21&  2014-09-18&   2002-06-16 &   SDSS&      181.39834&      49.472683&          0.082712&  GALAXY&                            &     ELG&     SN Ib/c\\
 Gaia14aaq&      168.98055&      49.38777&   20.07&   18.43&  2014-09-22&    2015-04-18 &  SDSS&      168.98048&      49.387738&           0.00026&    STAR&                            M1&       ALS&     unknown\\
 Gaia14abe&      221.40848&      49.54213&   20.56&   19.98&  2014-10-10&    2003-04-04 &  SDSS&      221.40845&      49.542121&          0.124505&  GALAXY&                             &       ELG&     unknown\\
 Gaia14abf&      202.89825&      43.49502&   19.32&   18.77&  2014-09-13&     2003-06-01 &  SDSS&      202.89809&      43.494994&          0.163195&  GALAXY&                              &      ALG&     unknown\\
 Gaia14abn&      192.44347&      52.23241&   20.09&   19.61&  2014-09-20&    2002-04-17 &  SDSS&      192.44312&      52.232371&          0.073223&  GALAXY&        STARBURST&      ELG&     unknown\\
\hline                                   
\end{tabular}
\end{table*}
%\end{landscape}

%\begin{landscape}
\begin{table*}
\tiny
\caption{Spectral information for GNT targets.}             % title of Table
\label{table:table2b}      
\centering                          
\begin{tabular}{llllllllllllllll}        
\hline
\hline    
Name  &  Gaia-Ra & Gaia-Dec & Gmag & Gmag-Peak & Peak-Date & Spec-Obsdate &  Spec-Tels & Spec-Ra & Spec-Dec & Redshift &  Class &  Subclass &  Type & Comments \\             
\hline     
  GNTJ000121.47-001140.29&     0.33945&    -0.19453&   19.89&   19.33&         2016-07-07&    2001-10-21&    SDSS&     0.339471&    -0.194535&          0.461516&  GALAXY&                         &                        low S/N&  Wrong z& \\
  GNTJ000426.46+160346.13&      1.11023&      16.06281&   19.06&   18.13&    2017-01-14&    2001-12-08& SDSS&      1.110209&      16.062791&          0.063098&  GALAXY&                       &                           E+A& \\
  GNTJ001019.55+025340.28&      2.58145&      2.89452&   19.19&   18.80&     2017-01-11&     2014-11-27& SDSS&      2.581443&      2.894539&           0.58875&     QSO&              BROADLINE&                          & \\
  GNTJ002026.66+334607.55&      5.11107&      33.76876&   19.30&   18.60&    2017-05-23&    2013-10-05&  SDSS&      5.111053&      33.768762&          0.123615&  GALAXY&                       &                            ALG& \\
  GNTJ002326.09+282112.86&      5.85873&      28.35357&   19.29&   18.34&    2017-05-24&    2015-12-05&  SDSS&      5.858735&      28.353559&          0.242617&     QSO&    STARBURST BROADLINE&          & \\
  GNTJ002742.67-003858.23&      6.92781&    -0.64951&   19.78&   19.46&       2016-12-12&     2005-12-30& SDSS&      6.927811&    -0.649497&          0.193833&  GALAXY&                       &                               ALG& \\
  GNTJ003217.08+185256.15&      8.07118&      18.88226&   20.35&   19.71&    2017-06-15&     2014-09-03& SDSS&      8.071209&      18.882252&          0.530666&     QSO&          AGN BROADLINE&                 & \\
  GNTJ003719.16+261312.23&      9.32982&      26.22006&   18.73&   18.01&    2017-05-26&     2013-01-17& SDSS&      9.329845&      26.220078&          0.147671&  GALAXY&              BROADLINE&                ELG&   Sey1.9&\\
  GNTJ004244.39+294134.73&      10.68496&      29.69298&   20.67&   20.05&  2016-11-30&     2015-12-06& SDSS&      10.684997&      29.692977&          0.306435&     QSO&    STARBURST BROADLINE&         & \\
  GNTJ005950.19+143648.25&      14.95912&      14.61340&   19.69&   19.17&  2016-11-29&     2000-10-04& SDSS&      14.959116&      14.613416&          0.187316&     QSO&    STARBURST BROADLINE&         & \\
 
\hline                                   
\end{tabular}
\end{table*}
\end{landscape}
\end{appendix}

\end{document}